\documentclass[alpha-refs]{wiley-article}
\usepackage{amsmath}
\usepackage{graphicx}%
\usepackage{amsfonts}%
\usepackage{amssymb}
\usepackage{pdflscape}
\usepackage{floatrow}
\usepackage{blindtext}

\RequirePackage[colorlinks,citecolor=blue,urlcolor=blue]{hyperref}
 \usepackage{natbib}
\usepackage{floatrow}
\usepackage{subcaption}
\allowdisplaybreaks

\usepackage{enumitem}
\usepackage{setspace}
\usepackage{caption}
\usepackage{blindtext}
\usepackage{multirow}
\newfloatcommand{capbtabbox}{table}[][\FBwidth]

\newcommand{\bbeta}{ \mbox{\boldmath $ \beta $} }

\newcommand{\bpsi}{ \mbox{\boldmath $\psi$} }

\newcommand{\bgamma}{ \mbox{\boldmath $\gamma$} }

\newcommand{\bzero}{\textbf{0}}
\newcommand{\bone}{\textbf{1}}

\newcommand{\bA}{\textbf{A}}

\newcommand{\bE}{\textbf{E}}

\newcommand{\bI}{\textbf{I}}

\newcommand{\bL}{\textbf{L}}

\newcommand{\bV}{\textbf{V}}

\newcommand{\bW}{\textbf{W}}
\newcommand{\bx}{\textbf{x}}
\newcommand{\bX}{\textbf{X}}
\newcommand{\by}{\textbf{y}}
\newcommand{\bY}{\textbf{Y}}

\papertype{Original Article}

\title{Pollution State Modeling for Mexico City}

\author[1\authfn{1}]{Philip A. White}%\thanks{paw27@stat.duke.edu}}
\author[1]{Alan E. Gelfand PhD}%\thanks{alan@stat.duke.edu}}
\author[2]{Eliane R. Rodrigues PhD}% \thanks{eliane@math.unam.mx}}
\author[3]{Guadalupe Tzintzun} %\thanks{guadalupe.tzintzun@inecc.gob.mx}}

\affil[1]{Department of Statistical Science, Duke University, Durham, NC, United States of America}
\affil[2]{Instituto de Matem\'{a}ticas,
Universidad Nacional Aut\'{o}noma de M\'{e}xico, Mexico}
\affil[3]{Instituto Nacional de Ecolog\'{\i}a y Cambio Clim\'{a}tico,
Secretar\'{\i}a de Medio Ambiente y Recursos Naturales, Mexico}

\contrib[\authfn{1}]{Corresponding Author}
\corraddress{Philip White, Department of Statistical Science, Duke University, Durham, NC, 27708, United States of America}
\corremail{philawhite@gmail.com}
\runningauthor{White et al.}

\fundinginfo{
The work of Eliane Rodrigues was partially funded by the project PAPIIT-IN102416 of the Direcci\'{o}n General de Apoyo al Personal Acad\'{e}mico of the Universidad Nacional Aut\'{o}noma de M\'{e}xico, Mexico (DGAPA-UNAM).
}

\begin{document}

\maketitle

\begin{abstract}
Ground-level ozone and particulate matter pollutants are associated with a variety of health issues and increased mortality. For this reason, Mexican environmental agencies regulate pollutant levels. In addition, Mexico City defines pollution emergencies using thresholds that rely on regional maxima for ozone and particulate matter with diameter less than 10 micrometers ($\text{PM}_{10}$). To predict local pollution emergencies and to assess compliance to Mexican ambient air quality standards, we analyze hourly ozone and $\text{PM}_{10}$ measurements from 24 stations across Mexico City from 2017 using a bivariate spatiotemporal model. Using this model, we predict future pollutant levels using current weather conditions and recent pollutant concentrations. Using hourly pollutant projections, we predict regional maxima needed to estimate the probability of future pollution emergencies. We discuss how predicted compliance to legislated pollution limits varies across regions within Mexico City in 2017. We find that predicted probability of pollution emergencies is limited to a few time periods. In contrast, we show that predicted exceedance of Mexican ambient air quality standards is a common, nearly daily occurrence.

\end{abstract}

\section{Introduction}\label{sec:intro}

Long-term exposure to air pollution is strongly linked with respiratory and cardiovascular disease and leads to increased mortality as well as hospital admissions \citep[see, e.g.,][]{brunekreef2002}. Particulate matter (PM) is defined to be solid particles and liquid droplets in the air. PM comes from direct emissions (primary particles) and chemical reactions between other pollutants (secondary particles). Particulate matter, and in particular PM with diameter less than 10 $\mu$m ($\text{PM}_{10}$), is known to increase human mortality and morbidity \citep[see, e.g.,][]{brunekreef2002,pope2006,loomis2013,hoek2013}. Because PM generally has a short lifetime, urban and other high emission areas generally have higher concentrations of $\text{PM}_{10}$ than rural areas \citep[see][as an example]{clements2012}.

Unlike PM, ground-level ozone ($O_3$) is not emitted directly but is instead formed by chemical reactions between of nitrogen oxides and volatile organic compounds, a reaction that requires heat and sunshine \citep[see, e.g., ][]{sillman1999}. Ozone is often as high in rural areas as it is in urban areas \citep[see, e.g., ][]{angle1989,sillman1999,duenas2004}. Ozone is linked to a variety of negative health outcomes, including short-term respiratory events, long-term respiratory disease, increased mortality, and low birth weight \citep[see, e.g.,][]{lippmann1989,salam2005,bell2006,weschler2006}. Because of these adverse outcomes, regulatory agencies institute policies to monitor and limit pollution levels, especially $\text{PM}_{10}$ and ozone. Urban areas are often monitored more closely to protect larger populations due to higher pollution levels found in urban environments \citep[see, e.g.,][]{heal2014}. The detrimental health effects of air pollution in the Mexico City metropolitan area are well-studied \citep[see][]{mage1996,romieu1996,hernandez1997,loomis1999,bravo2002,barraza2008,riojas2014}. Thus, Mexican authorities have implemented a variety of regulations to control pollution levels in Mexico, and specifically in Mexico City.

In spite of several polices implemented by environmental authorities in Mexico and Mexico City over the past 30 years, the city and its metropolitan area still suffer with high levels of pollution \citep[see, e.g.,][]{bravo2002,zavala2009,rodriguez2016,davis2017,INECC2017,gouveia2018}.
Some of the most recent measures implemented are new thresholds limiting ozone and $\text{PM}_{10}$ concentrations nation-wide which decreased allowable pollution levels relative to previous thresholds \citep{nom14a,nom14b}. Thresholds are updated every five years based on current research on the effect of pollutants on human health. In these new standards, the ozone thresholds were reduced to 95 parts per billion (ppb) or, equivalently, 0.095 parts per million (ppm) for \emph{hourly}
ozone and 70 ppb for \emph{eight-hour average} ozone \citep{nom14b}. Additionally, the allowable 24-hour average $\text{PM}_{10}$ concentration threshold was lowered to 75 micrograms per cubic meter ($\mu g/m^3$) \citep{nom14a}. These thresholds are not used to reduce pollution levels but are instead established to ensure human health protection and to evaluate air quality.

By comparison, the United States Environmental Protection Agency (EPA) limits 24-hour average $\text{PM}_{10}$ concentration to not exceed 150 $\mu g/m^3$ and 8-hour average ozone concentration to not exceed 70 ppb. \citep{CAA1990}. The European Union restricts 24-hour average $\text{PM}_{10}$ concentration to not exceed 50 $\mu g/m^3$ and 8-hour average ozone concentration to not exceed 120 ppb \citep{EU2016}. Thus, Mexican ambient air quality standards (which we denote MAAQS) are progressive when compared to American and European standards. Mexico City's pollution emergencies, however, are not related to the Mexican national standards and instead use more permissive thresholds. 

Mexico City's thresholds, established by the Atmospheric Environmental Contingency Program in Mexico City, are used to indicate times when pollutant concentrations are high enough to cause significant damage to human health \citep{mc2016}. Thus, the goals of Mexico City's Atmospheric Environmental Contingency Program differ from those specified for Mexico's ambient air quality standards. When emergency phases (or events) are activated, the aim is to control emission levels to decrease air pollution and its harmful effects to the population. It is worth mentioning that thresholds have decreased significantly. For instance, the thresholds for declaring the equivalent emergencies 1995-2000 were 1.5-2 times the current limits, depending on the type of emergency declared \citep{mc1996}.

Mexico City and its metropolitan area are split into five regions: northeast (NE), northwest (NW), central (CE), southeast (SE), and southwest (SW). Within these five regions, there are 24 monitoring stations that record both hourly ozone and $\text{PM}_{10}$ levels during the year 2017. To control the health risks associated with high ozone and $\text{PM}_{10}$, environmental alerts are declared if either hourly ozone or 24-hour average PM$_{10}$ levels exceed certain pollutant-specific thresholds which rely on regulatory suggestions that differ from those presented in \citet{nom14a,nom14b}.
Depending on the levels of the pollutant, either a phase I or a phase II alert is declared. Phase I is declared when hourly ozone exceeds $L_{1}^{O}=$0.154 ppm (154 ppb) or 24-hour average $\text{PM}_{10}$ exceeds $L_{1}^{PM}=$ 214 $\mu g/m^{3}$. During a phase I emergency, people are encouraged to limit outdoor time, exercise, smoking, and consumption of gas. Additionally, several transportation protocols are instituted to reduce vehicular emissions. Similarly, phase II is declared when hourly ozone exceeds $L_{2}^{O}=$ 0.204 ppm (204 ppb) or 24-hour average $\text{PM}_{10}$ exceeds $L_{2}^{PM}=$ 354 $\mu g/m^{3}$. Phase II institutes stricter protocols than phase I, including restricting circulation of official vehicles and strictly limiting civilian and commercial emissions. See \citet{mc2016} for details regarding Mexico City's pollution emergency phases.
%Depending on the pollutant's level either a phase I or a phase II alert is declared. Phase I is declared when ozone exceeds $L_{1}^{O}=$0.165 ppm (165 ppb) or $\text{PM}_{10}$ exceeds $L_{1}^{PM}=$ 220 $\mu g/m^{3}$. During a phase I emergency, people are encouraged to reduce time spent outdoor, excercise, smoking, consumption of gas \cite{mc2016}. Additionally, several transportation protocols are instituted to reduce vehicular emissions \cite{mc2016}. Similarly, phase II is declared when ozone exceeds thresholds $L_{2}^{O}=$ 0.220 ppm (220 ppb) or $\text{PM}_{10}$ exceeds $L_{2}^{PM}=$ 320 $\mu g/m^{3}$ (\textbf{I need a citation for this. The most recent legislation that I found on this gave different limits than the thresholds you have given me. Were they changed recently? If so, could I get a citation?}).

Compared to MAAQS, Mexico City's Atmospheric Environmental Contingency Program thresholds are more tolerant of high pollution levels. The phase I thresholds for ozone are 1.6 times the Mexican legal limits, while the phase I thresholds for $\text{PM}_{10}$ are almost three times MAAQS. The phase II thresholds are roughly two and five times MAAQS for ozone and $\text{PM}_{10}$, respectively. The protocols for phase I or phase II are the same regardless of the pollutant that triggered the alert. If ozone thresholds are exceeded in any region, i.e., the maximum over any station within the region, then emergency phases are declared city-wide (i.e., in all regions), where the phase is determined by which threshold ($L_{1}^{O}$ or $L_{2}^{O}$) was exceeded.

On the other hand, $\text{PM}_{10}$ exceedances could trigger regional or city-wide phase alerts, depending on which stations exceed the allowable limits. More explicitly, if the maximum 24-hour average over stations within the same region exceeds a $\text{PM}_{10}$ threshold, then the environmental alert is declared only in that region. However, if the maxima for two or more regions exceed a given $\text{PM}_{10}$ threshold, then the environmental alert is declared over the entire metropolitan area (i.e. in all five regions). This description is summarized in Table \ref{tab:phase}.

\begin{table}[H]
\centering
\footnotesize
\begin{tabular}{| l | l | l |}
  \hline
 Phase & Region-wide Alert & City-wide Alert \\
  \hline
 \rule{0pt}{2.5ex}None & $\bullet$ $\text{PM}_{10} < L_{1}^{PM}$ and $O_3 < L_{1}^{O}$ for all regions & $\bullet$ $\text{PM}_{10} < L_{1}^{PM}$ and $O_3 < L_{1}^{O}$ for all regions \\
   & $\bullet$ No higher-order alerts supersede &  \\
   \hline
 \rule{0pt}{2.5ex}I & $\bullet$ $\text{PM}_{10} \geq L_{1}^{PM}$ within the region &  $\bullet$ $ O_3 \geq L_{1}^{O}$ for any region \\
      & $\bullet$ And no higher-order alerts supersede  & $\bullet$ Or $\text{PM}_{10} \geq L_{1}^{PM}$ for two or more regions \\
    & & $\bullet$ And no higher-order alerts supersede \\
    \hline
 \rule{0pt}{2.5ex}II & $\bullet$ $\text{PM}_{10} \geq L_{2}^{PM}$ within the region &  $\bullet$ $ O_3 \geq L_{2}^{O}$ for any region \\
     & & $\bullet$ Or $\text{PM}_{20} \geq L_{2}^{PM}$ for two or more regions \\
   \hline
\end{tabular}
\caption{Description of Mexico City emergency phase alerts. Note that ozone thresholds are for hourly ozone, while $\text{PM}_{10}$ limits are for 24-hour running average $\text{PM}_{10}$. }\label{tab:phase}
\end{table}

Pollution emergency phases are only suspended when pollution levels for every station drop below phase I thresholds (i.e. the conditions for no phase alerts are met).  For practical reasons, evaluation of the emergency phases is carried out three times daily at 10 AM, 3 PM, and 8 PM \citep{mc2016}. Ultimately, however, phase activation and suspension are dependent on meteorological forecasts in addition to observed pollution levels. Because the additional meteorological criteria are not explicitly outlined, we do not attempt to predict actual phase occurrence but instead quantify the \emph{risk} of a phase occurrence.

The contribution here is to understand and predict how often Mexico City was at risk of a pollution emergency in terms of (1) the Atmospheric Environmental Contingency Program in Mexico City and (2) current Mexican ambient air quality standards. For both, we assess how the risk of dangerous pollution varies over city regions and over time. As described above, for Mexico City's Atmospheric Environmental Contingency Program, alerts are triggered when one or more stations exceeds thresholds.
%Depending on the pollutant, alerts can be limited to a specific region or declared over the whole metropolitan area.
Thus, emergency phases depend entirely upon pollutant maxima within each region. Furthermore, environmental alerts are often summarized over coarser temporal scales, like days, rather than the measurement level (hours) or the three hours of evaluation (10 AM, 3 PM, and 8 PM). So, daily emergency phases depend on pollutant maxima over hours of evaluation and stations within each region. Regarding MAAQS, we can do inference at each of three natural spatial scales: station-level, region-level, or city-level. Again, we may be interested in exceedances occurring on a daily scale rather than hourly. Therefore, we again need maxima over time (and potentially space depending on the spatial scale selected).

In summary, the foregoing tasks addressed here, the analyses of emergency contingency plan and Mexican ambient air quality standards, rely on the same pollution level data. Therefore, we develop, using model choice over a selection of models, a single hierarchical bivariate spatiotemporal model for hourly ozone and $\text{PM}_{10}$ levels. Predictions from our model serve two practical purposes: First, our predictions allow us to carry out probabilistic inference about pollution emergency states or national compliance issues. Second, if implemented in practice, our model could warn of potential pollution emergencies or compliance problems, allowing regional and city-wide adjustments and responses to be made earlier. From this model, all prediction and inference regarding emergency phases and legislation-based exceedances becomes a post-model fitting exercise, as we demonstrate. 

By now there is a rich literature on modeling both $O_{3}$ and $\text{PM}$ at both coarse ($10 \mu m$) and fine ($2.5 \mu m$) scale. Here, we highlight some examples relevant to our analysis. \citet{sahu2007} use a hierarchical space-time model to model square-root ozone with the goal of assessing long term trends in ozone in Ohio. \citet{cocchi2007} adopt a hierarchical model for log-$\text{PM}_{10}$ concentrations to characterize the effect of meteorological conditions on the $\text{PM}_{10}$ process and to estimate $\text{PM}_{10}$ at unmonitored locations. \citet{berrocal2010} model square-root ozone using data of two types, output from numerical models and data collected from monitoring networks, that are misaligned on spatial scales using spatially-varying regression coefficients \citep{gelfand2003}. \citet{huang2018} model log-$\text{PM}_{10}$ and log-nitrogen dioxide ($\text{NO}_2$) jointly and assessed the effect of these pollutants on health outcomes in Scotland. Similarly, we adopt a hierarchical space-time model with site-specific regression and auto-regressive coefficients for square-root ozone and log-$\text{PM}_{10}$ concentrations in Mexico City, Mexico.

In this paper, we start by presenting and discussing the Mexico City pollution dataset in Section \ref{sec:data}, highlighting data characteristics that inform modeling decisions. In Section \ref{sec:model}, we discuss modeling decisions, model fitting, inference, and  selection. We present and discuss results in the context of both Mexico City's Atmospheric Environmental Contingency Program and MAAQS in Section \ref{sec:res}. In this section, we first present a comprehensive analysis of Mexico City's phase alert system, predicting pollution levels and associated phase levels at 10 AM, 3 PM, and 8 PM to mirror the actual phase activation and suspension procedure. Then, we carry out inference on MAAQS exceedances and compare these results to emergency phase predictions to demonstrate differences between these standards. We provide concluding discussion regarding our results and statistical modeling in Section \ref{sec:conc}.

\section{Mexico City Pollution Dataset}\label{sec:data}

In this dataset, we have hourly ozone and $\text{PM}_{10}$ measurements at $N_s = 24$ stations across Mexico City, Mexico for the duration of 2017. Ozone and $\text{PM}_{10}$ measurements are obtained minute by minute at each station, and the hourly measurement reported is an average of the 60 minute-by-minute measurements.
%\textbf{I think we need more information. Is this an average over the previous hour???}
Let $Y^{O}_{it}$, $Y^{PM}_{it}$ denote ozone and $\text{PM}_{10}$ levels, respectively, at station $i$ and time $t$ with units of hours. Consequently, we observe measurements over $N_t = 8760$ times at each station, giving $N = 210240$ pairs of ozone and $\text{PM}_{10}$ concentration across the $24$ stations, across the entire year.\footnote{Missing hourly measurements were imputed using the corresponding measurements at the nearest station within the same region. If no stations in that region recorded a measurement at that time, then the nearest station in a different region provided the missing value. This was done prior to our receiving the data for analysis.} Relative humidity (RH) and temperature (TMP) are measured over the same space-time grid as ozone and $\text{PM}_{10}$ and are used as explanatory variables for both ozone and $\text{PM}_{10}$.

\begin{figure}[H]
  \begin{center}
      \includegraphics[width=.75\textwidth]{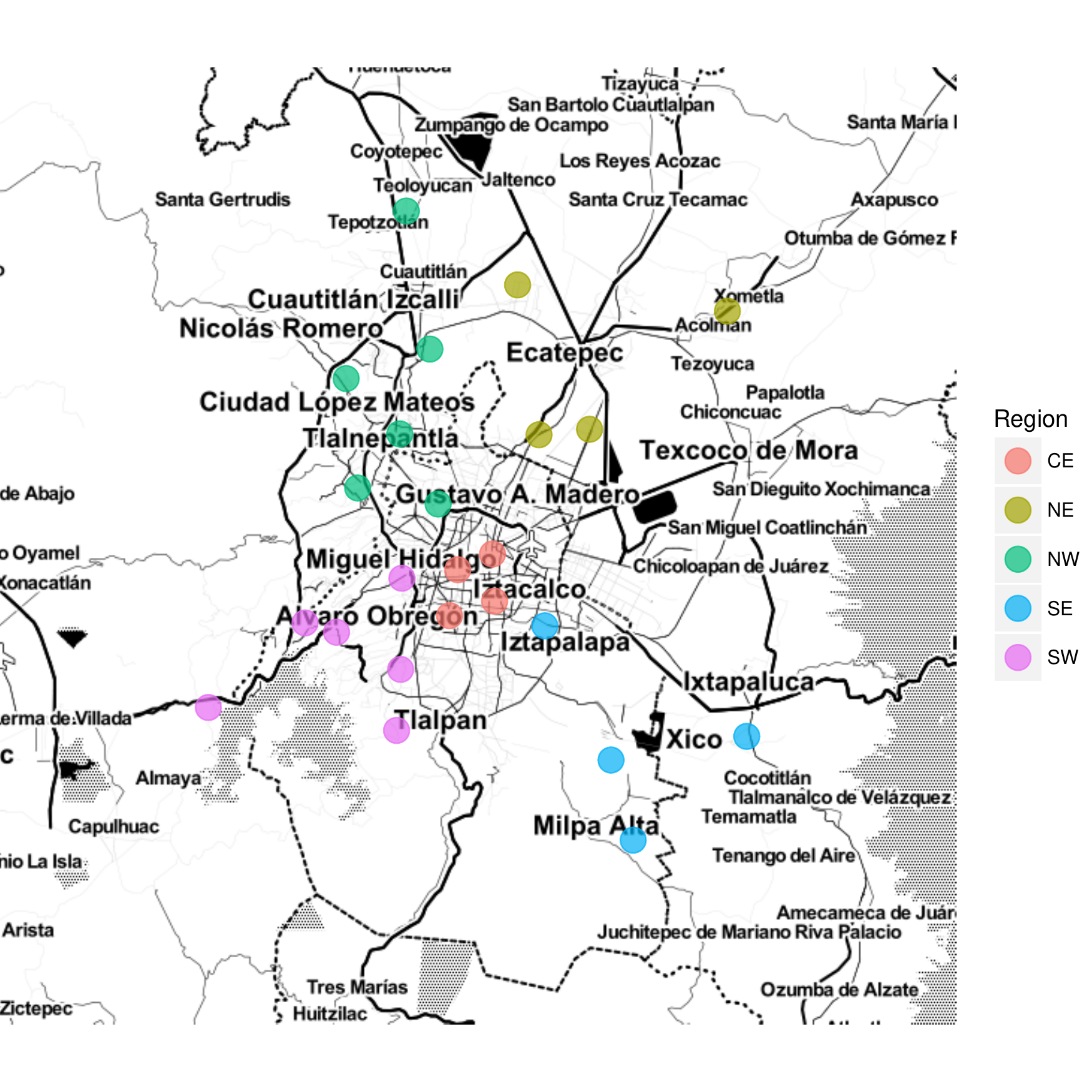}
  \end{center}
  \vspace{-12mm}
       \caption{Station locations with regional labels. %\textbf{It is quite clear that Acolman (ACO) -- see the top right seems to belong better to the NE region, but perhaps I don't understand how these regions are defined.}
}\label{fig:station_loc}
\end{figure}

As mentioned above, Mexico City is partitioned into five regions which are employed for defining environmental alert phases (See Section \ref{sec:intro}). Abbreviated station names, corresponding regions, and annual summaries of pollution levels are given in Table \ref{tab:stat}. Station locations are plotted in Figure \ref{fig:station_loc} using the R package GGMAP \citep{kahle2013}. Besides being on the main wind path (from NE to SW) and therefore receiving many ozone precursors from the NE region, the CE region is heavily-trafficked by automobile. The SW region, located and the end of the NE-SW wind corridor, receives ozone produced along this wind path, and this ozone stays trapped in the SW region due to mountains on its southwest boundary.
 \vspace{-6mm}
\vspace{-3mm}
\begin{table}[H]
\centering
\footnotesize
\begin{tabular}{rclcc}
  \hline
   &  &  & Annual & Annual \\
Region & Stations & Station Names &  Average Ozone & Average $\text{PM}_{10}$ \\
  \hline
Northeast & 4 & ACO, SAG, VIF, XAL & 28 ppb & 59 $\mu g / m^3$  \\
Northwest & 6 & ATI, CAM, CUT, & 27 ppb &49  $\mu g / m^3$  \\
 &  &  FAC, TLA, TLI &  &  \\
Central & 4 & BJU, HGM, IZT, MER & 29 ppb  &44 $\mu g / m^3$  \\
Southeast & 4 &  CHO, MPA,TAH, UIZ  & 36 ppb & 45 $\mu g / m^3$  \\
Southwest & 6 & AJM, CUA, INN & 34 ppb &33 $\mu g / m^3$  \\
 &  &  MGH, PED, SFE &  & \\
   \hline
\end{tabular}
\caption{Station names and regions. Average ozone and $\text{PM}_{10}$ across regions are given.}\label{tab:stat}
\end{table}
  \vspace{-2mm}
Note that the number of stations in each region differs.
%Consequently, the induced distribution of the regional maximum differs, where regions with more stations are more likely to trigger emergency phases or to exceed Mexico's legislated thresholds if all regions have similar pollution levels, all else being equal.
Moreover, pollution levels appear to vary over the regions. The northeast region has the highest average $\text{PM}_{10}$, while the southeast and southwest regions have the highest average ozone.   Hence, the regional maxima, used for phase alerts, are expected to have very different hourly distributions.  Note that, with the maxima being taken over a small number of stations in each region, there is no reason to attempt to employ extreme value theory here.  We model at the station-level rather than at the regional level so that the regional maximum distributions are \emph{induced} by the station-level modeling.  In this regard, because ozone and $\text{PM}_{10}$ are strictly non-negative, we consider modeling the station data using either transformations to $\mathbb{R}$ or using strictly positive data models.

\begin{figure}[H]
  \begin{center}
   \begin{subfigure}[b]{.32\textwidth}
      \includegraphics[width=\textwidth]{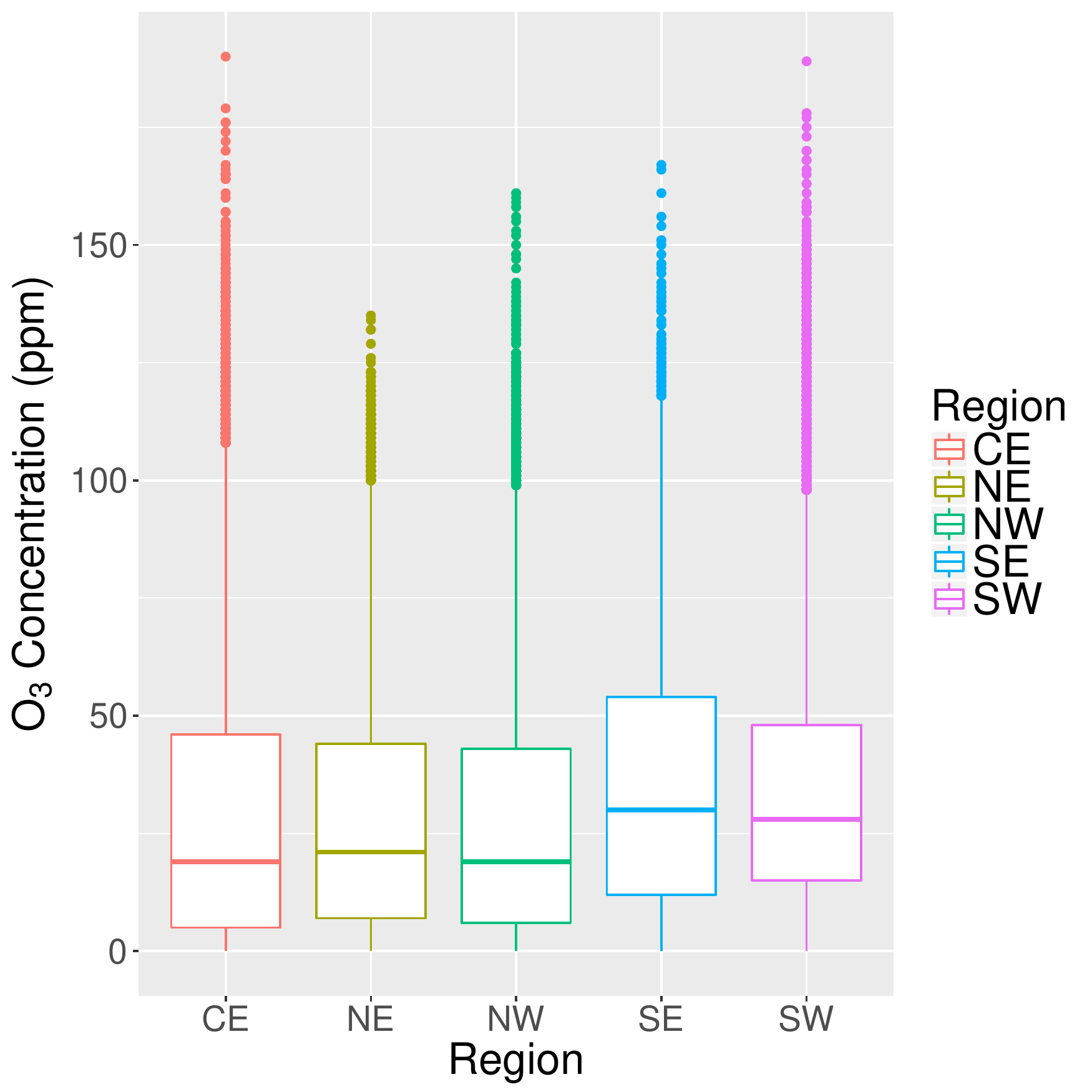}
      \subcaption{Region $\text{O}_{3}$ boxplot}\label{fig:reg_O3_plots}
   \end{subfigure}
     \begin{subfigure}[b]{.32\textwidth}
     \includegraphics[width=\textwidth]{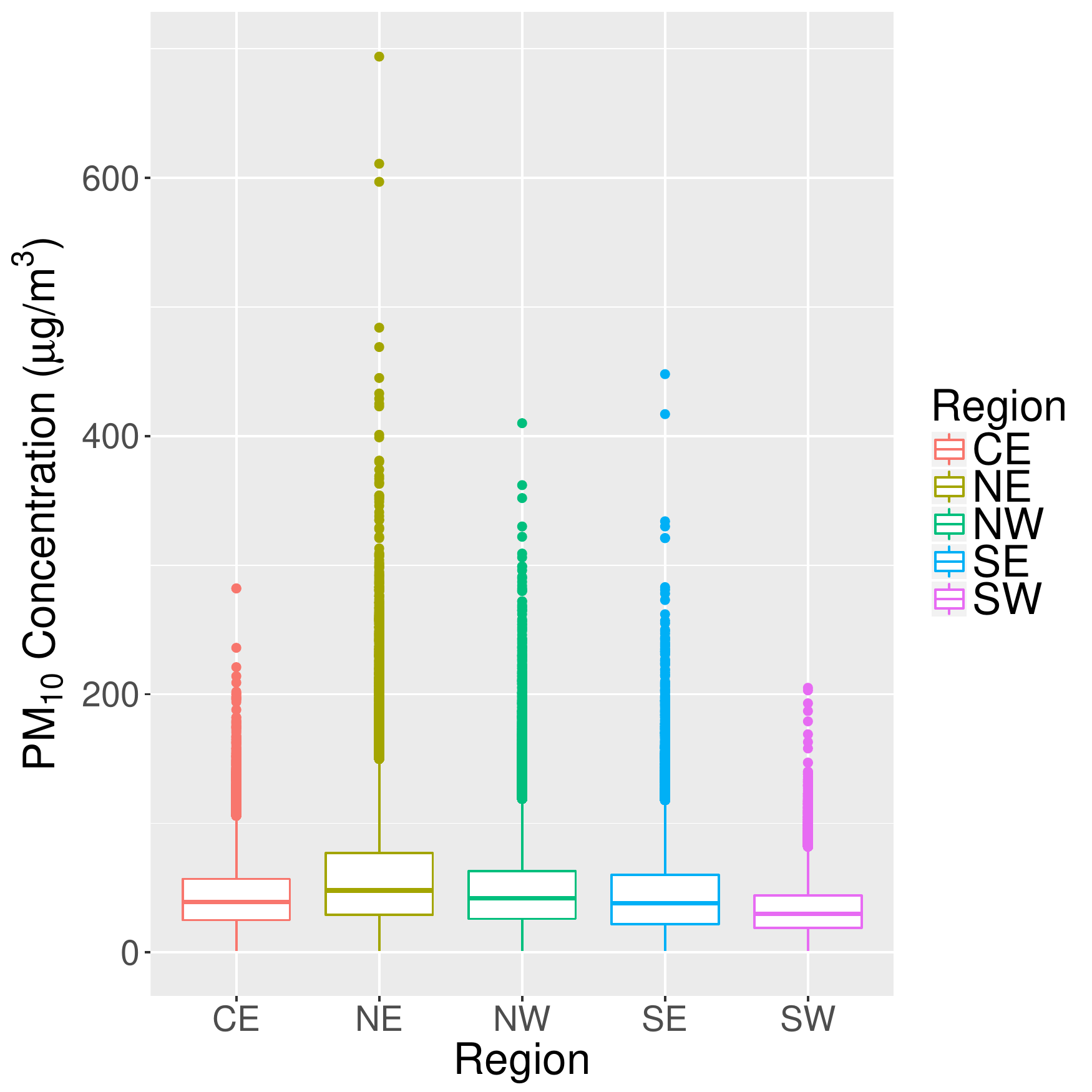}
           \subcaption{Regional $\text{PM}_{10}$ boxplot}\label{fig:reg_pm_plots}
   \end{subfigure}
        \begin{subfigure}[b]{.32\textwidth}
     \includegraphics[width=\textwidth]{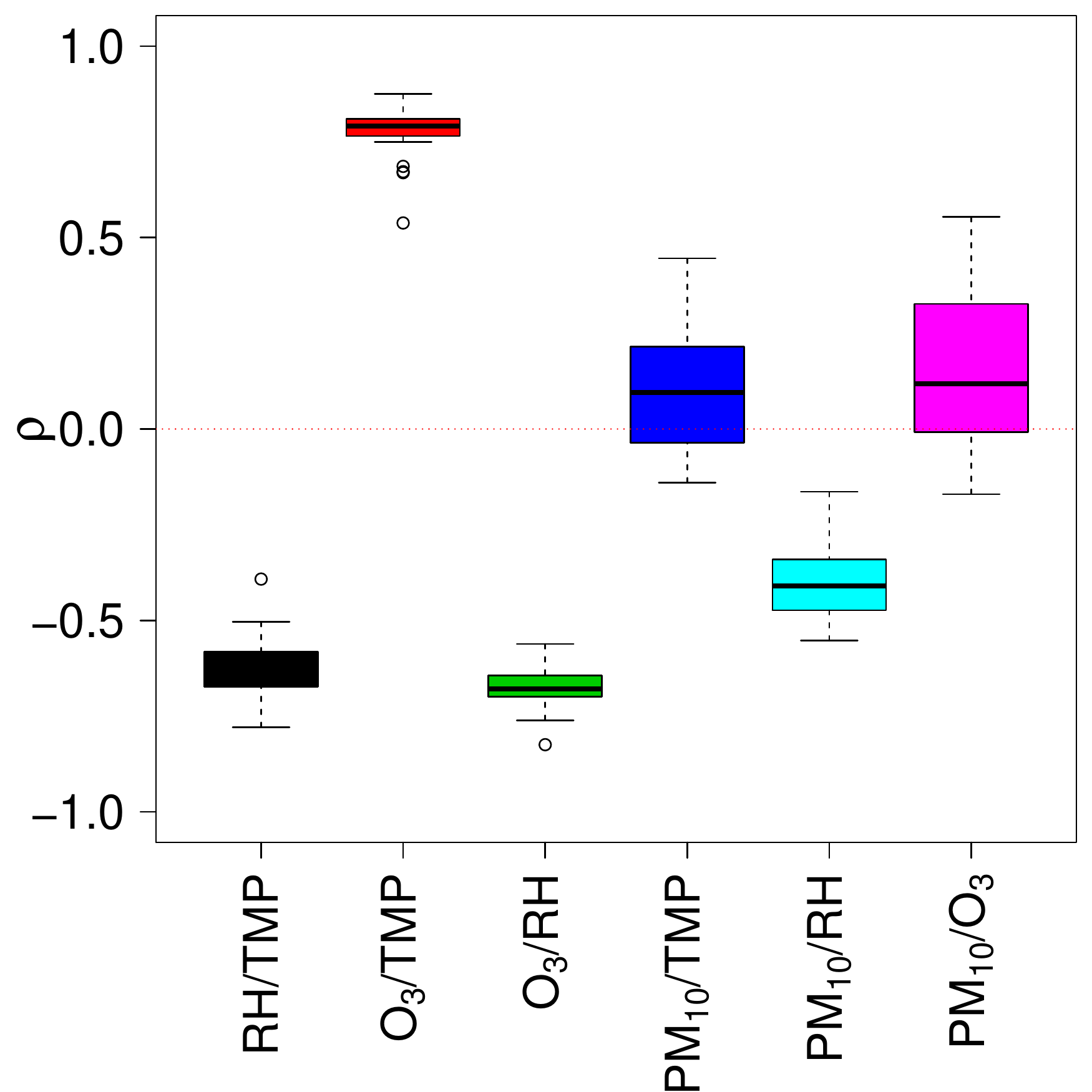}
           \subcaption{Correlation boxplots}\label{fig:correlation_plots}
   \end{subfigure}
  \end{center}
  \vspace{-4mm}
       \caption{Region-specific boxplots for (Left) ozone and (Center) $\text{PM}_{10}$. (Right) Site-specific Spearman's $\rho$ for (from left to right) relative humidity and temperature, ozone and temperature, ozone and relative humidity, $\text{PM}_{10}$ and temperature, $\text{PM}_{10}$ and relative humidity, and $\text{PM}_{10}$ and ozone. }\label{fig:reg_box_plot}
\end{figure}

Region-specific box-plots for ozone and $\text{PM}_{10}$ are plotted in Figures \ref{fig:reg_O3_plots} and \ref{fig:reg_pm_plots}. For ozone, the SE region has the highest mean, but the CE and SW regions have the most extreme values. Note that the NE region has the highest average $\text{PM}_{10}$, as well as the most extreme values. This is because the NE region houses a large industrial section that generates many direct pollutants, including particulate matter. To explore the relationships between covariates (RH and TMP), outcomes (ozone and $\text{PM}_{10}$), and covariates and outcomes, we compute the station-specific Spearman's $\rho$ for all covariate-covariate, covariate-outcome, and outcome-outcome relationships. As a rank correlation, Spearman's $\rho$ avoids concern regarding transformations and outlying values. We plot these site-specific correlation coefficients in Figure \ref{fig:correlation_plots}.
\vspace{-2mm}
The relationships between ozone and covariates (RH and TMP) are strong for all sites, while the relationships between $\text{PM}_{10}$ and covariates (RH and TMP) vary much more across sites.  However, there appears to be a strong negative correlation between RH and $\text{PM}_{10}$ for all locations. On the other hand, Spearman's $\rho$ varies greatly across sites for TMP and $\text{PM}_{10}$. Similarly, there appear to be important relationships between ozone and $\text{PM}_{10}$ depending on the site, motivating the use of a joint model for ozone and $\text{PM}_{10}$. The variability of the outcome, covariate, and outcome-covariate relationships across stations motivates a hierarchical model for covariate effects.

%\textbf{Alan: Increase axis label size on most plots}
%\vspace{-2mm}
%\begin{figure}[H]
%  \begin{center}
%      \includegraphics[width=.5\textwidth]{xy_cor}
%  \end{center}
%  \vspace{-4mm}
%       \caption{Site-specific Spearman's $\rho$ for (Far Left) relative humidity and temperature (Left) ozone and temperature (Left-middle) ozone and relative humidity (Right-middle) $\text{PM}_{10}$ and temperature (Right) $\text{PM}_{10}$ and relative humidity (Far Right) $\text{PM}_{10}$ and ozone.}\label{fig:correlation_plots}
%\end{figure}

We find strong daily and weekly patterns for both pollutants. Using residual analysis for a model with site-specific effects for meteorological covariates, we still observe strong seasonal patterns for both day and week. Additionally, our preliminary analyses reveal strong correlation between the variance of residuals and the mean for both pollutants. This correlation could be addressed through modeling in a variety of ways. First, and most simply, one could use a variance stabilizing transformation (VST) to address the correlation between the mean and variance (e.g. log, square-root, Box-Cox) as was done by, for example, \citet{sahu2007,cocchi2007,berrocal2010,huang2018}. Alternatively, we could use heteroscedastic models that specify variance directly as a function of hour or month. We consider both modeling approaches (transformations and heteroscedasticity) in Section \ref{sec:mod_comp}. This exploratory analysis is provided in the online supplement.

\section{Methods and Models}\label{sec:model}

Given the exploratory analysis, a time-series analysis is certainly warranted. Because the data are collected hourly, because the exposure standards are at the scale of hours (or functions of hours), and because we can identify useful discrete lags which are difficult to capture with covariance specifications, we elect to work with discrete time rather than continuous time.
%and prediction at a sub-hourly scale is not of interest, we argue that modeling the data in discrete time is natural and preferable.
%However, we briefly discuss how similar continuous-time models could be posed and fit.
Additionally, our exploratory analysis suggests that a model using either a VST or time-varying variance may describe the data more accurately than models using non-transformed data or homoscedastic models. As a result, we envision the model for these data to be
\begin{align}\label{eq:mod3a}
Y^{O}_{it} &= \bx_{i(t-1)}^T \bbeta_{1i} + {\bL^{O}_{it}}^T\bgamma_{1i}+ \psi_{1i} + \epsilon_{1it} \\
Y^{PM}_{it} &= \bx_{i(t-1)}^T \bbeta_{2i}+ {\bL^{PM}_{it}}^T\bgamma_{2i} + \psi_{2i} + \epsilon_{2it} , \nonumber
\end{align}
where $Y^{O}_{it}$ is ozone concentration (or square-root ozone) and $Y^{PM}_{it}$ is $\text{PM}_{10}$ concentration (or log-$\text{PM}_{10}$) at site $i$ and hour $t$. 
%The log-scale may be used to match the support of the data, otherwise a likelihood with positive support should be used. 
Here, $\bx_{i(t-1)}$ includes an intercept, temperature, and relative humidity at site $i$ and time $t-1$. We use covariates from the previous hour because one of the primary purposes of this model is one-hour-ahead predictions for pollutants and corresponding phase alerts and exceedance probabilities and $\bx_{it}$ will not be available for such prediction.
%Is model-based imputation interesting or should we just stick to how they are going to impute missing values?

The parameters $\bbeta =( \bbeta_{11},...,\bbeta_{1N_s},\bbeta_{21},...,\bbeta_{2N_s})$ are station-specific regression coefficients for both $\text{PM}_{10}$ and ozone. Because we imagine that the effect of humidity on pollutant concentrations is similar from region to region, we model regression coefficients exchangeably and hierarchically, centering effects on respective common means \citep[see][for introductory thoughts on such hierarchical modeling]{gelman2014}. We define ${\bL^{O}_{it}}$ and ${\bL^{PM}_{it}}$ to be generic vectors of the lagged observations for ozone and $\text{PM}_{10}$, respectively with $\bgamma =( \bgamma_{11},...,\bgamma_{1N_s},\bgamma_{21},...,\bgamma_{2N_s})$ as corresponding site-specific autoregressive coefficients. Lags for observations in early January 2017 (${\bL^{O}_{it}}$ and ${\bL^{PM}_{it}}$) may depend upon observations from December 2016. The choice of components of ${\bL^{O}_{it}}$ and ${\bL^{PM}_{it}}$ becomes the model choice issue which we take up in Section 3.3. As with $\bbeta$, we model $\bgamma$ hierarchically. Then, we have pure error terms, $\epsilon_{1it} \overset{iid}{\sim} N(0,\sigma^2_1)$ and $\epsilon_{2it} \overset{iid}{\sim} N(0,\sigma^2_2)$, or $\epsilon_{1it} \overset{ind}{\sim} N(0,\sigma^2_{1t})$ and $\epsilon_{2it} \overset{ind}{\sim} N(0,\sigma^2_{2t})$ for the heteroscedastic formulation.
%More generally, a heteroscedastic model assumes $\epsilon_{1i}(t) \overset{ind}{\sim} N(0,\sigma^2_{1t})$ and $\epsilon_{2i}(t) \overset{ind}{\sim} N(0,\sigma^2_{2t})$.

Finally, to bring in spatial structure across the sites, jointly, $\psi_{1i},\psi_{2i}$ follow a bivariate conditionally autoregressive (CAR) model using coregionalization of two independent CAR models $V_{1i}$ and $V_{2i}$ \citep[see][]{rue2005,banerjee2014}. Coregionalization allows flexible, multivariate modeling \citep[see, e.g.,][]{matheron1982,grzebyk1994,wackernagel1994,banerjee2014}. Explicitly,
\begin{align*}
\begin{pmatrix}
\psi_{1i} \\ \psi_{2i} \end{pmatrix}  &= \bA_\psi (V_{1i} , V_{2i})^T \\
\bA_\psi &= \begin{pmatrix}
a^{(\psi)}_{11} & 0 \\ a^{(\psi)}_{12} & a^{(\psi)}_{22}, \end{pmatrix}
\end{align*}
where $\bV_1 = (V_{11},V_{12},...,V_{1N_s})^T$ and $\bV_2= (V_{21},V_{22},...,V_{2N_s})^T$. Equivalently, we can view $a^{(\psi)}_{11}$ and $a^{(\psi)}_{22}$ as scale parameters for the $\bV_1$ and $\bV_2$, a fact we use in model fitting (See Appendix \ref{app:gibbs3}).  %\textbf{Alan:  Help!  How can the $V$'s have a scale (or variance) parameter and we also have an associated scale parameter in the $A$ matrix?  You can not identify both?  Ahh, I now see that you clarified this below but it would probably be cleaner to remove the $\tau^2$'s and have an unrestricted $A$ matrix.  You certainly do not need expression (3) below.}
Because there are not natural borders or edges shared between stations, it is natural that $\bV_1$ and $\bV_2$ would use an inverse distance-dependent proximity matrix which we denote $\bW$. We assume the same distant-dependent CAR structure for both $\bV_1$ and $\bV_2$, where weights are proportional to $\exp(-ad)$ with $d$ denoting the distance between locations and $a$ being the inverse of the maximum distance between stations. 
%\textbf{Alan: How do we know that $e^{-d}$ provides the appropriate scaling for distances? Why now $e^{-ad}$ for a suitable $a$?}  
For common proximity matrix $\bW$, if we let $D_W$ be diagonal with ${(D_W)}_{ii} = w_{i+}$, where $w_{i+} =  \sum_i \bW_{ij}$, then, the (unscaled) precision matrix of $\bV_1$ and $\bV_2$ is $Q=D_W-W$.

\subsection{Priors, Model Fitting, and Prediction}\label{sec:priors}

We model regression and autoregressive coefficients $\bbeta_{1i}$, $\bbeta_{2i}$, $\bgamma_{1i}$, and $\bgamma_{2i}$ hierarchically,
\begin{equation}
\begin{aligned}[c]
\bbeta_{1i} &\sim N(\bbeta_{01} , \Sigma_{\beta_1} ), \\
\bbeta_{2i} &\sim N(\bbeta_{02} ,\Sigma_{\beta_2} ), \\
\bbeta_{01} &\sim N(\bzero, 10^3 \, \bI ), \\
\bbeta_{02} &\sim N(\bzero, 10^3 \, \bI ), \\
\end{aligned}
\hspace{10mm}
\begin{aligned}[c]
\bgamma_{1i} &\sim N(\bgamma_{01} , \Sigma_{\gamma_1} ), \\
\bgamma_{2i} &\sim N(\bgamma_{02} ,\Sigma_{\gamma_2} ), \\
\bgamma_{01} &\sim N(\bzero, 10^3 \, \bI), \\
\bgamma_{02} &\sim N(\bzero, 10^3 \, \bI), \\
\end{aligned}
\hspace{10mm}
\begin{aligned}[c]
\Sigma_{\beta_1} &\sim IW(10^3 \, \bI ,p + 1), \\
\Sigma_{\beta_2} &\sim IW(10^3 \, \bI , p + 1 ), \\
\Sigma_{\gamma_1} &\sim IW(10^3 \, \bI ,n_{1l} + 1), \\
\Sigma_{\gamma_2} &\sim IW(10^3 \, \bI , n_{2l} + 1 ), \\
\end{aligned}
\end{equation}
where $p = 3$ is the number of regressors including the intercept, $n_{1l}$ is the number of lags for ozone, and $n_{2l}$ is the number of lags for $\text{PM}_{10}$. By this, we assume that station-specific regression and autoregression coefficients are exchangeable. For the variance terms in the likelihood and the CAR prior, we assume that
\begin{equation}
\begin{aligned}[c]
\sigma_1^2 &\sim IG(1,1), \\
\sigma_2^2 &\sim IG(1,1), \\
\end{aligned}
\hspace{20mm}
\begin{aligned}[c]
{a^{(\psi)}_{11}}^2 &\sim IG(1,1), \\
{a^{(\psi)}_{22}}^2  &\sim IG(1,1). \\
\end{aligned}
\end{equation}
Lastly, we assume ${a^{(\psi)}_{12}}^2 \sim N(0,10^3)$. Model fitting details via a Gibbs sampler are given in Appendices \ref{app:gibbs3} and \ref{app:gibbs4}. This model could be fit sequentially, but this would require us to update model parameters 8760 times for each step an MCMC sampler, once for each hour in 2017.
%\textbf{Again, why introduce symbols just to set them =1?}
%For this problem, one of the primary modeling questions is which and how many lagged terms we should include.
%\textbf{This seems a bit silly.  You fit the model in the most efficient way, then you do inference!  I think the possible confusion here is the \textrm{repeated} fitting of the model according to the inference you wish to make.  However, with any set of hourly measurements up to a given time, the model is always fitted the same way.  And, that is what is in Appendices A and B. Also, for clarity, we we work with inference based upon 24-hour averages, I assume that you take 23 hours as observed and only predict the last one in order to create the 24-hour average?}

Prediction can be done in two ways, each with a different purpose: one predicts at unobserved values within the time range of the data (missing data) or one can predict future observations (forecasting). The former is viewed as \emph{retrospective} prediction, filling in missing data over sites and times. The latter is viewed as prospective, predicting the next hour given the data up to the current hour. We are interested in MAAQS exceedances for specific months on an hourly scale. In this case, we only train model parameters on data observed prior to our predictions. Instead of a fully sequential model fitting where the model is updated hourly, we fit the model up to the last hour of the previous month. This model is then used to predict for pollutant levels for the upcoming month, making all prediction in this setting prospective. When carrying out inference for all days simultaneously, we fit the model to all the data once. For pollution and phase predictions, we limit our prediction to 10 AM, 3 PM, and 8 PM, each day to match the times of phase activation and suspension. Even though we make one-hour-ahead predictions, our predictions depend on model parameters that are trained using all the data; thus, our phase analysis is, in a sense, retrospective even though predictions are prospective. This model allows us to make probabilistic inference about reaching the conditions for environmental phase alerts in Mexico City and about Mexico City's compliance with MAAQS. 
%\textbf{I think you might move the discussion of the previous paragraph here to illustrate when you will be retro- and when you will be pro- with regard to inference.  Then you could begin separate paragraphs for how you implement each case.}

In the missing data context, we suppose that arbitrary $Y^{O}_{it}$ or $Y^{PM}_{it}$ is unobserved. This could be due to limited sampling or for model validation on a holdout dataset, but we only take this predictive approach when comparing models in Section \ref{sec:mod_comp}. Each held-out observation is updated or imputed as a part of model fitting using a Gibbs sampler (See Appendix \ref{app:full_cond_pred} for details). 
%\textbf{This will confuse the reader since, in Section 2 we say that the data arrived with imputation of some sort already done. So, there is really no missing data unless we do holdout.  } 

To predict future pollution measurements using our model, Equation \ref{eq:mod3a}, we use the following formula:
\begin{align}\label{eq:mod3a_one}
Y^{O}_{i(t+1)} &= \bx_{it}^T \bbeta_{1i} + {\bL^O_{i(t+1)}}^T\bgamma_{1i}+ \psi_{1i}  + \epsilon_{1i(t+1)} \\
Y^{PM}_{i(t+1)} &= \bx_{it}^T \bbeta_{2i}+ {\bL^{PM}_{i(t+1)}}^T\bgamma_{2i} + \psi_{2i}  + \epsilon_{2i(t+1)}.\nonumber
\end{align}
Note that one-step-ahead predictions do not rely on future covariates. We use this type of prediction for both inferential tasks (See Sections \ref{sec:retro_phase} and \ref{sec:prospective}). 

In our setting, predicted phase alerts come from the one-hour-ahead ozone predictions ($\hat{Y}_{it}^O$) and the predicted 24-hour average $\text{PM}_{10}$ concentration ($\widehat{\overline{Y}_{it}^{PM}}$), where $\widehat{\overline{Y}_{it}^{PM}}$ is the average of the 23 most recent observed $\text{PM}_{10}$ concentrations ($Y_{i(t-1)}^{PM},...,Y_{i(t-23)}^{PM}$) and the forecasted $\text{PM}_{10}$ level ($\hat{Y}_{it}^{PM}$). Nationally legislated thresholds depend on 24-hour average $\text{PM}_{10}$ and on 8-hour average $O_3$. Similar to $\widehat{\overline{Y}_{it}^{PM}}$, predicted 8-hour average ozone concentration ($\widehat{\overline{Y}_{it}^{O}}$) is an average of a one-hour-ahead prediction and the previous seven ozone measurements ($Y_{i(t-1)}^{O},...,Y_{i(t-7)}^{O}$). For predictions of both $\widehat{\overline{Y}_{it}^{PM}}$ and $\widehat{\overline{Y}_{it}^{O}}$ on January 1, 2017,  we rely on hourly observations from December 31, 2016.

\subsection{Posterior Inference}\label{sec:post_analysis}

The primary inferential goal for this dataset is to assess how often the Mexico City metropolitan area (1) is at risk for declaring phase I or II emergencies and (2) exceeds MAAQS. For each task, we take different modeling approaches, as discussed in Section \ref{sec:priors}. To analyze the risk of phase I and II emergencies, we fit the model to all the data. In contrast, when examining pollution level exceedances, we fit the model sequentially. For both tasks, we use one-step-ahead predictions for ozone and $\text{PM}_{10}$ concentrations.
%\textbf{It seems like you will have already said this in the previous subsection.  Moreover, when will hsave said something about 24-hour averages or do you plan to skip them?  Ahh, this shows up below but, with regard to prediction, you still need to clarify how you are obtaining them.  Moreover, $d$ is one of the $365$ days of the year while 24-hour averages are not connected to days.  Altogether, I think the reader will be confused by all of this.}
These posterior predictions allow us to carry out probabilistic inference on emergency phases and MAAQS exceedances to assess how often the Mexico City metropolitan area was at risk of a pollution emergency and how often pollution levels were unsafe according to Mexican federal guidelines \citep{nom14a,nom14b}.
%Because we forecast, our inference and uncertainty quantification is retrospective and not forward looking. Instead, our goal is to assess how often the Mexico City metropolitan area was at risk of a pollution emergency and how often pollution levels were unsafe according to Mexican federal guidelines \citep{nom14a,nom14b}.

To define useful quantities, let $j$ index region and $d$ index day, such that each station $i \in j$ and each hour $t \in d$. Additionally, we define $\overline{Y_{it}}^{PM}$ to be the 24-hour running average of $\text{PM}_{10}$ at time $t$ and station $i$.
%\textbf{Eliane thought it made more sense to use $\text{d}$ and $\text{j}$ instead of $d$ and $j$. I can update this.}\textbf{Alan: Nope! $j$ and $d$ are not vectors.} 
We define the following maxima:
\begin{equation}\label{eq:max}
\begin{aligned}[c]
Z_{jt}^{O} &= \max_{i \in j} Y_{it}^{O}  \\
Z_{jt}^{PM} &= \max_{i \in j} \overline{Y}_{it}^{PM}
\end{aligned}
\hspace{10mm}
\begin{aligned}[c]
W_{jd}^{O} &= \max_{t \in d} \max_{i \in j} Y_{it}^{O} \\
W_{jd}^{PM} &= \max_{t \in d}\max_{i \in j} \overline{Y}_{it}^{PM},
\end{aligned}
\end{equation}
where $Z$'s are regional maxima for any hour $t$ and $W$'s are daily regional maxima. It is important to clarify that although these maxima often rely on data observed prior to time $t$ or day $d$, these quantities are used to define exceedances at time $t$ or day $d$.
There is limited literature about the distributions and properties of maxima for correlated random variables \citep[see][]{gupta1985,ho1996}. However, these examples are too constrained for our application. In the spatial literature, modeling extreme values, and sometimes maxima, is well-studied \citep[see, e.g.,][]{sang2009,sang2010,davison2012}; however, these approaches generally invoke generalized extreme value (GEV) distribution models. As noted in Section 2, our inference depends on relatively few maxima over few sites or hours, so GEV theory is not applicable. In fact, using the definitions in (\ref{eq:max}), we do not model the maxima directly.  Instead, we obtain the derived posterior predictive distribution for $Z_{jt}^{O}$, $Z_{jt}^{PM}$, $W_{jd}^{O}$, and $W_{jd}^{PM}$ from posterior predictive samples of $Y_{it}^{O}$ and $Y_{it}^{PM}$.

The states of Mexico City's phase alert system $S_{jt} \in \{0,1,2\}$ are completely determined by $Z_{jt}^{O}$ and $Z_{jt}^{PM}$ (see Section \ref{sec:intro} and Table \ref{tab:phase}). To obtain the maximum phase alert for a day $d$ in some region $j$ ($\max_{t \in d} S_{jt}$), we use $W_{jd}^{O}$ and $W_{jd}^{PM}$. One may also wish to infer the distribution of the highest phase alert in any region on day $d$ ($ \max_{j} \max_{t \in d} S_{jt}$). All these derived posterior quantities can be obtained after model fitting. Using derived posterior predictive distributions for various maxima, as well as associated phase states and threshold exceedances, we can compute hourly and daily probabilities of (possible) phase alerts and pollution exceedances regionally and city-wide. The utility of these probabilities is the insight they can provide regarding how often the Mexico City metropolitan area is at risk of a pollution emergency, even if phase alerts were not enacted due to meteorological forecasts.

Inference for the first task, analysis of the phase emergencies, requires analysis of regional pollution levels at 10 AM, 3 PM, and 8 PM. Specifically, phase states depend on maxima of stations over regions. If we carry out inference on a daily scale, double maxima are needed, maxima over hours and stations within regions. Because phase alerts are based upon one-hour ozone measurements and 24-hour average $\text{PM}_{10}$ \citep{mc2016}, we predict these averages as described in Section \ref{sec:priors}. These predictions allow us to compute predictions for derived quantities $\widehat{Z_{jt}^{O}}$, $ \widehat{Z_{jt}^{PM}}$, $\widehat{W_{jd}^{O}}$, and $ \widehat{W_{jd}^{PM}}$, which in turn define phase state predictions $\widehat{S_{jt}}$. Again, inference on phase predictions is our primary goal.

%\textbf{Add thoughts about the one-step-ahead prediction. How high do we need to observe to trigger a PM10 emergency or to exceed legislation. The same goes for ozone in the context of legislation; however, phase alerts depend on one-hour ozone. }
%We formalize the description of Table \ref{tab:phase} in Equation \ref{eq:formal_phase}.
%\begin{equation}\label{eq:formal_phase}
%S_{jt} =\begin{cases}
%   0 & \text{if }  \{ Z^{PM}_{jt} < L_1^{PM}  \} \cap \{  Z^{O}_{j't} < L_1^{O} \; \forall \;  j' \} \cap \{ \text{at most one } Z^{PM}_{j't} \geq L_1^{PM} \} \\
%   1 & \text{if }  \{ Z^{PM}_{jt} \geq L_1^{PM}  \} \cup \{ \text{if } \exists \; j' \; st \; \ Z^{O}_{j't} \geq L_1^{O} \cap  Z^{O}_{j*t} < L_1^{O} \; \forall \;  j* \neq j' \} \cup   \\
%   & \{ Z^{PM}_{j't} \geq L_1^{PM} \text{ for }  \} \\
%   2 & \text{if } x < 0 \\
%  \end{cases}
%\end{equation}

We also carry out similar inference on MAAQS exceedance for ozone and $\text{PM}_{10}$. Again, we are interested in regional (i.e. stations within a specified region) and city-level (i.e. at any station in the city) exceedance, hourly and daily. Similar, but not identical to phase alerts, inference for pollution exceedances relies upon maxima of one-hour ozone, eight-hour average ozone $\overline{Y_{it}}^{O}$, and 24-hour average $\text{PM}_{10}$. For eight-hour average ozone, we define $Z_{jt}^{\overline{O}}$ to be the regional maxima at time $t$ and $W_{jd}^{\overline{O}}$ to be the daily maxima for region $j$.
%\textbf{Again, some words are needed as to how this eight-hour average is forecasted.} 
In this case, thresholds are much lower than the thresholds Atmospheric Environmental Contingency Program in Mexico City \citep{nom14a,nom14b,mc2016}. Because nationally legislated ozone and $\text{PM}_{10}$ thresholds were specified to avoid reaching unsafe pollution levels, comparisons to MAAQS indicate how often Mexico City reaches unsafe pollution levels without triggering any city protocols. Such comparisons highlight important differences in how pollution emergencies are defined in Mexico City relative to nationally legislated levels.

\subsection{Model Selection}\label{sec:mod_comp}

In this subsection, we describe the model selection process leading to the model under which we carry out all the inference described in the previous subsections. Our model selection decision centers around answering how many and which lagged terms should be used in our spatiotemporal model. In our exploratory analyses, we argued that the variance of ozone and $\text{PM}_{10}$ vary with the time of day and time of year. We indicated that this could be remedied in one of two ways: (1) using variance-stabilizing transformations to address correlation between mean and variance in the data or (2) modeling the heteroscedasticity directly. For transformation approaches, we consider modeling the data on different scales (truncated, log, and square root) to stabilize the mean-variance correlation. To answer these modeling questions, we hold out 10\% of both pollutants and treat these as missing data.  Specifically, the locations and times of the hold-out data are selected at random, and both ozone and $\text{PM}_{10}$ are held-out at these location-time pairs so that model comparison can be made using joint predictions. We make predictions at these held-out observations and compare competing models based on several criteria: predictive mean squared error $(E(Y_i|\bY_{obs}) - y_i)^2$ (PMSE) or mean absolute error $|E(Y_i|\bY_{obs}) - y_i|$ (PMAE), $100\times \alpha$ \%  prediction interval coverage, and continuous rank probability scores (CRPS) \citep{gneiting2007}, where
\begin{equation}
\text{CRPS}(F_i,y_i) = \int^\infty_{-\infty} (F_i(x) -  \bone(x \geq y_i) )^2 dx = \bE|Y_i - y_i | - \frac{1}{2}\bE | Y_i - Y_{i'} | .
\end{equation}
Because we are utilizing MCMC to fit our model, we use posterior predictive samples for a Monte Carlo approximation of CRPS using an empirical CDF approximation \citep[see, e.g., ][]{kruger2016},
\begin{equation}
\text{CRPS}(\hat{F}^\text{ECDF}_i,y_i) = \frac{1}{M} \sum_{j=1}^M |Y_j - y_i| - \frac{1}{2M^2} \sum_{j=1}^M \sum_{k=1}^M | Y_j - Y_k| ,
\end{equation}
where $M$ is the number of MCMC samples used, $Y_i$ are predictions, and $y_i$ are observed values. We then average $\text{CRPS}(\hat{F}^\text{ECDF}_i,y_i)$ over all held-out data. In addition to being a proper scoring rule \citep{gneiting2007}, because CRPS considers how well the entire predictive distribution matches the observed data rather than only the predictive mean (MAE and MSE) or quantiles (prediction interval coverage), we prefer it as selection criterion. For multivariate predictions, as we have in this analysis, we consider the energy score (ES), which is a multivariate generalization of CRPS. For a set of multivariate predictions $\bY$, ES is defined as
\begin{equation}
\text{ES}(P,\by) = \frac{1}{2} E_P \left\| \bY - \bY' \right\|^\beta - E_P \left\| \bY - \by \right\|^\beta,
\end{equation}
where $\by$ is an observation, $\beta \in (0,2)$, and $P$ is a probability measure \citep{gneiting2007}. It is common to fix $\beta = 1$ \citep[see, e.g.,][]{gneiting2008,jordan2017}. For a set of $M$ MCMC predictions $ \bY = \bY_1,...,\bY_M$ for a held-out observation $\by$, the empirical ES reduces to
\begin{equation}\label{eq:es}
\text{ES}(\bY,\by) = \frac{1}{M} \sum^M_{j=1} \left\| \bY_j - \by \right\|  - \frac{1}{2M^2} \sum^M_{i=1} \sum^M_{j=1} \left\| \bY_i - \bY_j \right\|,
\end{equation}
as was discussed in \citet{gneiting2008}. Energy scores are scale-sensitive, meaning that if one of the variables has a much larger scale than other of the variables, it dominates the norms in Equation \ref{eq:es}. In our data, $\text{PM}_{10}$ concentration in $\mu g/m^3$ takes values larger than ozone in ppb. To assure that predictions for each pollutant are similarly weighted, we standardize the predictions and hold-out values for each pollutant (i.e. subtract the sample mean and divide by the sample standard deviation). Like CRPS, we average ES over all held-out data.

Interestingly, the heteroscedastic models with variance that varies over hour of the day and month of the year performed uniformly worse than homoscedastic counterparts that used VST's to stabilize the mean-variance correlation. Testing various combinations of square-root transformations, log transformations, and truncated distributions, we found that models using the square-root transformation for ozone and the log transformation for $\text{PM}_{10}$ gave the best predictive performance. So, for model selection, we only give the results for six models which use the square-root transformation for ozone and the log-transformation for $\text{PM}_{10}$ but differ in terms of which lags are included in the model. The results of this comparison are given in Table \ref{tab:mod_comp}.

\vspace{-3mm}
\begin{table}[H]
%\begin{longtable}{rrrrrrrrrrrr}
\centering
\footnotesize
\begin{tabular}{lllrrrrrrrrrr}
 & &  &  &  &  &  &  &  & & \\
 \hline
 & &   & $O_3$ & $O_3$  & $O_3$ & $O_3$ & $\text{PM}_{10}$ & $\text{PM}_{10}$  & $\text{PM}_{10}$ & $\text{PM}_{10}$ \\
  & Lags & ES & CRPS & RMSE  & MAE & Cov & CRPS & RMSE  & MAE & Cov \\
  \hline
%1 & TN  & (1,2) & 2.58 & 5.00 & 3.34 & 0.90 & 6.90 & 14.21 & 8.58 & 0.93 \\
%2 & TN  & (1,24) & 3.10 & 5.61 & 3.88 & 0.96 & 6.85 & 14.05 & 8.55 & 0.93 \\
%3 & TN & (1,2,24) & 2.57 & 4.96 & 3.34 & 0.90 & 6.85 & 14.03 & 8.54 & 0.93 \\
%4 & TN  & (1,2,12,24) & 2.57 & 4.97 & 3.35 & 0.90 & 6.86 & 14.05 & 8.56 & 0.93 \\
%5 & TN & (1:24) & 2.56 & 4.92 & 3.32 & 0.91 & 6.79 & 13.90 & 8.47 & 0.93 \\
%6 & TN & (1,24,168) & 3.08 & 5.57 & 3.86 & 0.96 & 6.85 & 14.02 & 8.55 & 0.93 \\
%7 & TN & (1,2,24,168) & 2.56 & 4.94 & 3.33 & 0.90 & 6.84 & 14.01 & 8.54 &  0.93 \\
%8 & TN& (1,2,12,24,168) & 2.57 & 4.95 & 3.34 & 0.90 & 6.85 & 14.01 & 8.56 & 0.93 \\
%9 & TN & (1:24,168) & 2.58 & 4.95 & 3.35 & 0.91 & 6.78 & 13.88 & 8.47 & 0.93 \\
%%%%%%%%%%%%%%%%%%%%%%%%%%%%%%%%%%
%9 & LN  & (1,2) & 3.48 & 5.81 & 3.88 & 0.90 & 6.68 & 14.52 & 8.72 & 0.92 \\
%10 & LN  & (1,24) & 3.85 & 5.66 & 3.91 & 0.94 & 6.57 & 14.12 & 8.58 & 0.92 \\
%11 & LN  & (1,2,24) & 3.35 & 5.39 & 3.63 & 0.91 & 6.57 & 14.13 & 8.58 & 0.92 \\
%12 & LN  & (1,2,12,24) & 3.36 & 5.51 & 3.73 & 0.91 & 6.56 & 14.10 & 8.62 & 0.92 \\
%13 & LN  & (1:24) & 3.33 & 5.46 & 3.72 & 0.91 & 6.50 & 13.95 & 8.49 & 0.92 \\
%14 & LN & (1,24,168) & 3.80 & 5.61 & 3.88 & 0.94 & 6.54 & 14.06 & 8.55 & 0.92 \\
%15 & LN & (1,2,24,168) & 3.35 & 5.36 & 3.62 & 0.91 & 6.55 & 14.08 & 8.60 &  0.92 \\
%16 & LN& (1,2,12,24,168) & 3.35 & 5.34 & 3.61 & 0.91 & 6.55 & 14.08 & 8.60 & 0.92 \\
%18 & LN & (1:24,168) & 3.36 & 5.63 & 3.83 & 0.91 & 6.49 & 13.91 & 8.50 & 0.92 \\
%%%%%%%%%%%%%%%%%%%%%%%%%%%%%%%%%%%
1 &  (1,2) & 0.2552 & 2.5392 & 5.0448 & 3.3409 & 0.8867 & 6.7263 & 14.5427 & 8.7725 & 0.9228 \\
2  & (1,2,24)& 0.2513 & 2.5158 & 4.9709 & 3.3035 & 0.8925 & 6.6176 & 14.1469 & 8.6189 & \textbf{0.9217} \\
\textbf{3}  & (1,2,24,168)&\textbf{0.2505} & \textbf{2.5140} & \textbf{4.9614} & \textbf{3.2982} & \textbf{0.8941} & \textbf{6.5947} & \textbf{14.0922} & \textbf{8.6285} &  0.9229 \\
4 &  (1,2,12) & 0.2540 & 2.5298 & 5.0274 & 3.3244 & 0.8887 & 6.6959 & 14.4314 & 8.7864 & 0.9230 \\
5  & (1,2,12,24)& 0.2509 & 2.5168 & 4.9726 & 3.3115 & 0.8917 & 6.6035 & 14.0981 & 8.6296 & 0.9220 \\
6  & (1,2,12,24,168)& 0.2507 & 2.5154 & 4.9651 & 3.2993 & \textbf{0.8941} & 6.5976 & 14.0972 & 8.6378 &  0.9225 \\
   \hline
\end{tabular}
\caption{Predictive model comparison. The ``Lags'' label indicates which lags are used for both outcomes. ``ES,'' ``CRPS,'' ``MSE,'' ``MAE,'' and ``Cov'' head columns giving ES, CRPS, MSE, MAE, and 90\% prediction interval coverage. Best performances are indicated with bold text.
%\textbf{Models with dashes haven't finished running, except 36. This model had major stability issues. These models are really slow with missing data because the lags change with imputation, so this doesn't scale super well. }
}\label{tab:mod_comp}
\end{table}

We further note that in preliminary modeling, we found that models which included a lag-3 and other higher order lags or that excluded lag-2 saw no improvement in terms of prediction; thus, we arrived at the models included in Table \ref{tab:mod_comp}. Given these results, we argue that the best model for ozone and $\text{PM}_{10}$ uses lags 1, 2, 24, and 168.   So, the ensuing results are presented for this model.

\section{Results and Discussion}\label{sec:res}

We present our inference based on a joint model for ozone and $\text{PM}_{10}$ with four lags (1, 2, 24, and 168). We use a Gibbs sampler to obtain 100,000 posterior samples after a burn-in of 10,000 iterations. Posterior parameter inference is discussed in the online supplement, and these results validate many of the modeling decisions suggested by our exploratory analysis in Section \ref{sec:data} and discussed in Section \ref{sec:model}. Because we have $N = 210240$ observations, the predictive space is large ($2 \times N \approx 4 \times 10^5$). Thus, we thin the posterior predictive samples using every 10$^\text{th}$ sample. By thinning, we make 10,000 roughly independent predictions. These predictions are used to carry out analyses in Sections \ref{sec:retro_phase} and \ref{sec:prospective}.

\subsection{Analysis of the Phase Alert System}\label{sec:retro_phase}

\begin{figure}[H]
  \begin{center}
    \hspace{.1\textwidth}
  \includegraphics[width=.6\textwidth]{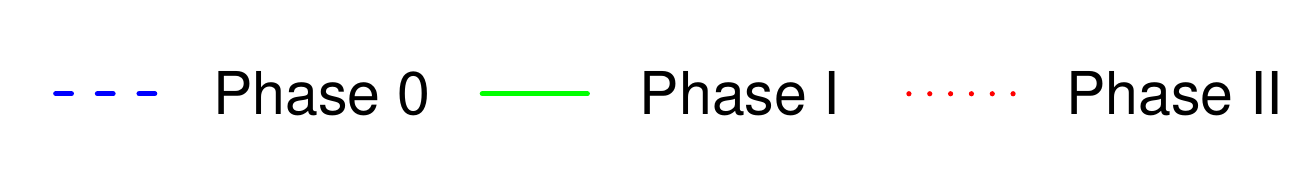}

  \vspace{-6mm}
      \includegraphics[width=\textwidth]{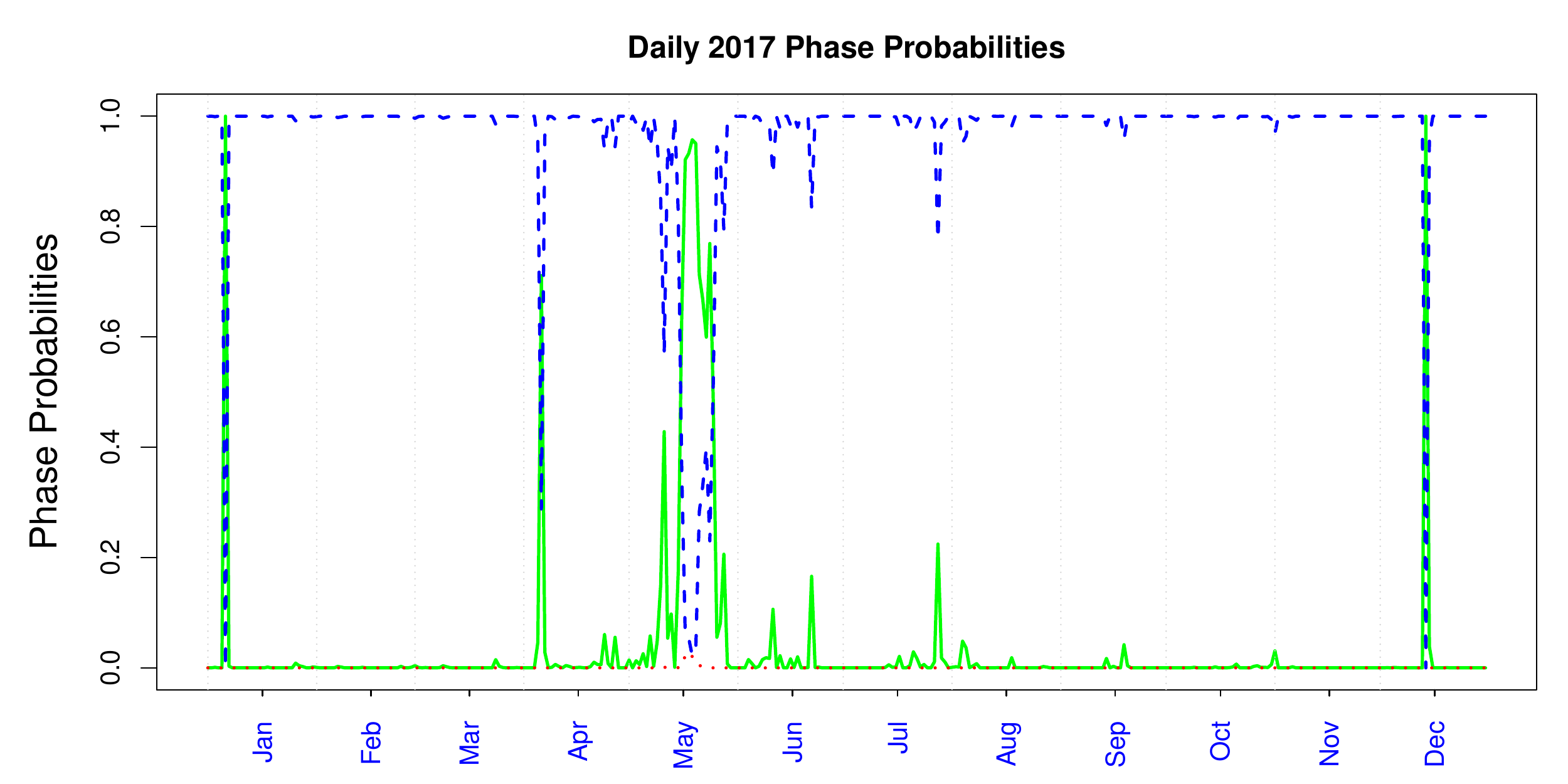}
  \end{center}
  \vspace{-8mm}
       \caption{Phase probabilities in Mexico City, aggregated over all regions.}\label{fig:phase_all}
\end{figure}

\begin{figure}[H]
  \begin{center}
    \hspace{.1\textwidth}
  \includegraphics[width=.6\textwidth]{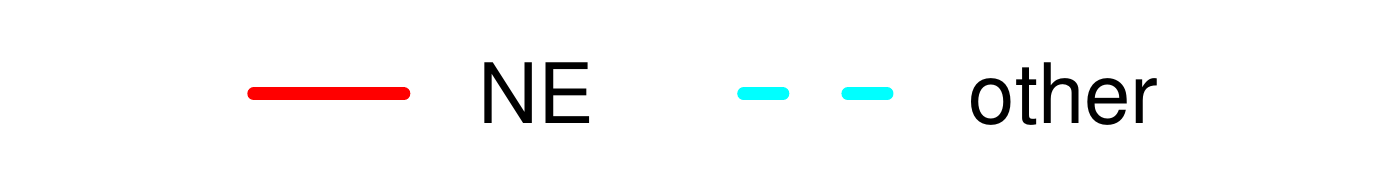}

  \vspace{-6mm}

      \includegraphics[width=\textwidth]{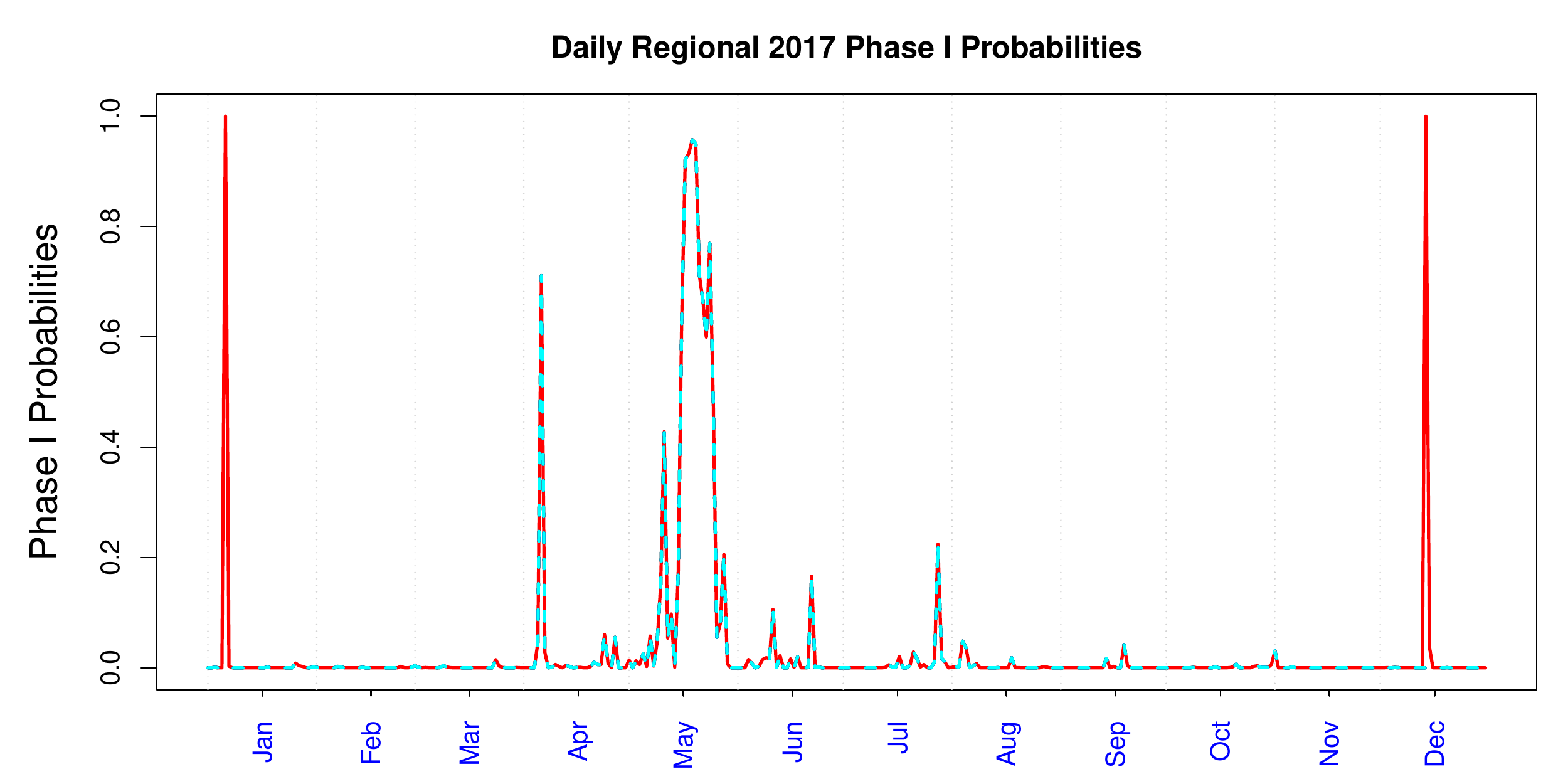}
  \end{center}
  \vspace{-8mm}
       \caption{Daily phase I probabilities for Mexico City over the year by region. Phase II probabilities are not included because they are uniformly low.}\label{fig:phase_region}
\end{figure}

In this section, we analyze Mexico City's phase alert system to identify when the Mexico City metropolitan area was predicted to be at risk for pollution emergencies. For this, we use one-hour-ahead predictions for pollution levels each day at the three decision times reference in Section \ref{sec:intro}: 10 AM, 3 PM, and 8 PM. Thus, our analysis predicts at three hours per day, altogether 1095 hours in 2017. This allows us to assess probabilities of the risk of phase alerts given the most recent weather conditions and pollution levels. Again, we note that the risk of a phase alert is not the same as a phase alert. As discussed above, we use parameter values trained on the entire dataset which enables effective prediction in early months. Because the pollutant thresholds for triggering phase alerts are very high, most of the year has very low probabilities for phase activation. In May of 2017, however, Mexico City was featured prominently in the news for having dangerously high ozone levels which led to an activation of a phase I pollution emergency. Phase probabilities aggregated over regions ($P(\max_j S_{jd} =k)$ for state $k$) are displayed in Figure \ref{fig:phase_all}. Regional phase I probabilities for each day ($P(S_{jd} = k)$ for phase $k$) are given in Figure \ref{fig:phase_region}. In Figure \ref{fig:phase_region}, we do not show phase II probabilities because they are so low. Additionally, we only display region NE compared to other regions because all other regions overlap (See Figure \ref{fig:phase_region}). Both plots (Figures \ref{fig:phase_all} and \ref{fig:phase_region}) show high probabilities ($>1/2$) of phase I activation from May 16th to May 25, coinciding with the time of the actually declared phase I emergency. Because this phase I alert was triggered by ozone levels, the emergency was declared city-wide, as indicated by the agreement of regional curves in Figure \ref{fig:phase_region}. 
\vspace{-6mm}
\vspace{-12mm}
\vspace{-3mm}
On April 6th, predicted ozone levels were sufficient to trigger a phase I emergency city-wide, although a phase alert was not declared. On only two occasions, one in January and one in December, was any region at risk of activating the emergency contingency plan due to $\text{PM}_{10}$ levels. These high phase I probabilities were limited to the northeast region (See the red peaks in Figure \ref{fig:phase_region}). Again, it is worth noting that a phase I alert triggered by $\text{PM}_{10}$ corresponds to $\text{PM}_{10}$ levels that are nearly three times the levels specified as safe by Mexican legislation \citep{nom14b}.
%\textbf{Why do I see only 2 colors (NE and SW) in this fig? And, why do we need a horizontal line at .5? Perhaps, separate panels for each region would look better?}

In Table \ref{tab:phase_region}, we provide posterior means and 95\% credible intervals for the number of hours and days for which the Mexico City metropolitan area is at risk for pollution emergencies ($\sum_d \bone(S_{jd} = k) $ for phase $k$). The first thing to notice is that there are very few hours and days when the metropolitan area or its sub-regions are at risk of pollution emergencies. Note that the posterior predictive mean for risk of a pollution emergency is 11 days and 11 hours for the central, northwest, southeast, and southwest regions. These counts are not necessarily reflective of conditions in central and northwest regions. Instead, these counts are indicative of predicted city-wide phase I alerts due to predicted ozone exceedances in the southeast and southwest regions, one in April and 10 in May (see Figures \ref{fig:phase_all} and \ref{fig:phase_region}). The northeast region is the only region that had more average predicted hours and days of pollution emergency than other regions. We predict six hours of risk for phase I emergencies in the northeast region due to $\text{PM}_{10}$ levels over two non-consecutive days, one day in January and one in December. Because the northeast region was the only region where a predicted phase alert was triggered by $\text{PM}_{10}$, the predicted risk of a phase alert was limited to the northeast region. No phase alert was declared even though predicted phase probabilities were equal to one. Thus, the reason for not declaring an emergency must be attributed to meteorological conditions. While we do know the exact rationale for not declaring a phase emergency, we speculate that the emergency was not declared because these predicted phase risks were transient, lasting only one day each.

\begin{table}[H]
\centering
\small
\begin{tabular}{|l|rrrrrr|}
\hline
 &\multicolumn{6}{c|}{ \textbf{Hours (total of 1095 || 3 hours / day) }} \\
 & CE & NE & NW & SE & SW & Any  \\
  \hline
%No Phase & $1084 \pm 4$ & $1078 \pm  4$  & $1084 \pm  4$  & $1084 \pm  4$  & $1084 \pm  4$  & $1078 \pm 4$  \\
%Phase I & $11 \pm 4$ & $17 \pm 4$ & $11 \pm 4$ & $11 \pm 4$ & $11 \pm 4$ & $17 \pm 4$  \\
%Phase II & 0.09 (0,1)& 0.09 (0,1)& 0.09 (0,1)& 0.09 (0,1) & 0.09 (0,1)& 0.09 (0,1)\\
No Phase & $1084 \pm 4$ & $1078 \pm  4$  & $1084 \pm  4$  & $1084 \pm  4$  & $1084 \pm  4$  & $1078 \pm 4$  \\
Phase I & $11 \pm 4$ & $17 \pm 4$ & $11 \pm 4$ & $11 \pm 4$ & $11 \pm 4$ & $17 \pm 4$  \\
Phase II & 0.09 (0,1)& 0.09 (0,1)& 0.09 (0,1)& 0.09 (0,1) & 0.09 (0,1)& 0.09 (0,1)\\
   \hline
 &\multicolumn{6}{c|}{\textbf{Days (total of 365)} } \\
No Phase & $354 \pm 4$ & $352 \pm  4$  & $354 \pm  4$  & $354 \pm  4$  & $354 \pm  4$  & $352 \pm 4$  \\
Phase I & $11 \pm 4$ & $13 \pm 4$ & $11 \pm 4$ & $11 \pm 4$ & $11 \pm 4$ & $13 \pm 4$  \\
Phase II & 0.09 (0,1)& 0.09 (0,1)& 0.09 (0,1)& 0.09 (0,1) & 0.09 (0,1)& 0.09 (0,1)\\
   \hline
\end{tabular}
\caption{One-hour-ahead posterior predictive estimates for the (Top) Number of hours in each phase state for each region (Bottom) Number of days for which that phase state was attained (the maxima attained each day). Posterior means and 95\% credible intervals are given for each region, using $\pm$ or parentheses. The ``Any'' label indicates that this is the maximum across regions. }\label{tab:phase_region}
\end{table}

\subsection{Comparison of Mexico City to Mexican Legislated Thresholds}\label{sec:prospective}

In this section, we examine the probability that maxima within regions exceed MAAQS on a given day ($W_{jd}^O$ and $W_{jd}^{PM}$ from Section \ref{sec:post_analysis}). In contrast to phase alert probabilities, which are generally very low, exceedance probabilities are often high through much of the year. Because MAAQS are more reflective of healthy levels of ozone and $\text{PM}_{10}$, comparison between the exceedance probabilities and emergency phase probabilities highlights how often Mexico City has harmful pollution levels without triggering phase alerts. Additionally, this analysis gives insight into the probability of triggering phase alerts in Mexico City if MAAQS were adopted for Mexico City's Atmospheric Environmental Contingency Program. For our purposes, we group either type of ozone exceedance, one or eight-hour, together. In the online supplement, we focus on three months, April, August, and December, to illustrate how exceedance probabilities change over the course of the year.

\begin{figure}[H]
  \begin{center}
  \includegraphics[width=.6\textwidth]{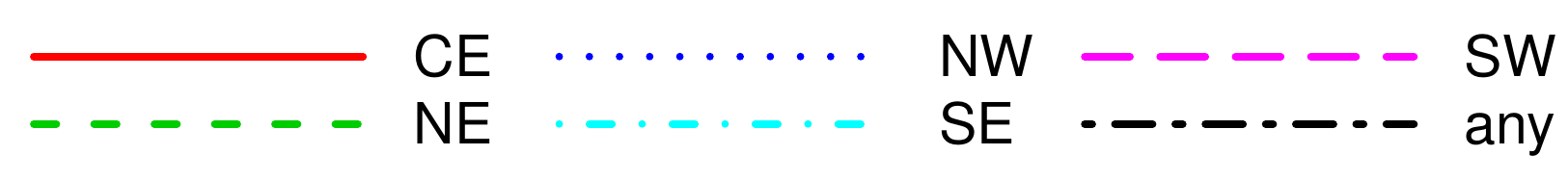}
  
 \vspace{-2mm}
 
   \begin{subfigure}[b]{.32\textwidth}
      \includegraphics[width=\textwidth]{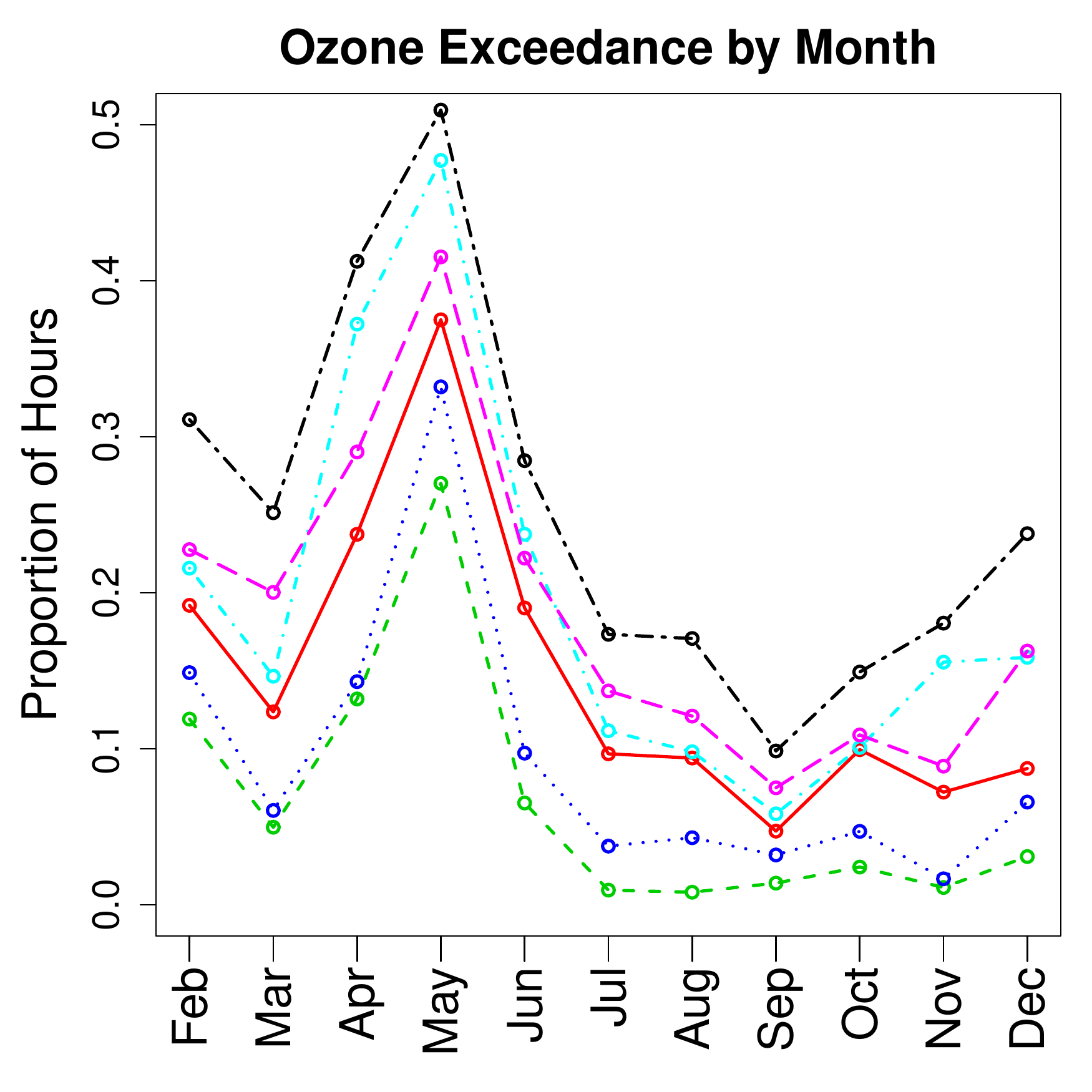}
      \subcaption{$\text{O}_{3}$ exceedance by month}
   \end{subfigure}
      \begin{subfigure}[b]{.32\textwidth}
      \includegraphics[width=\textwidth]{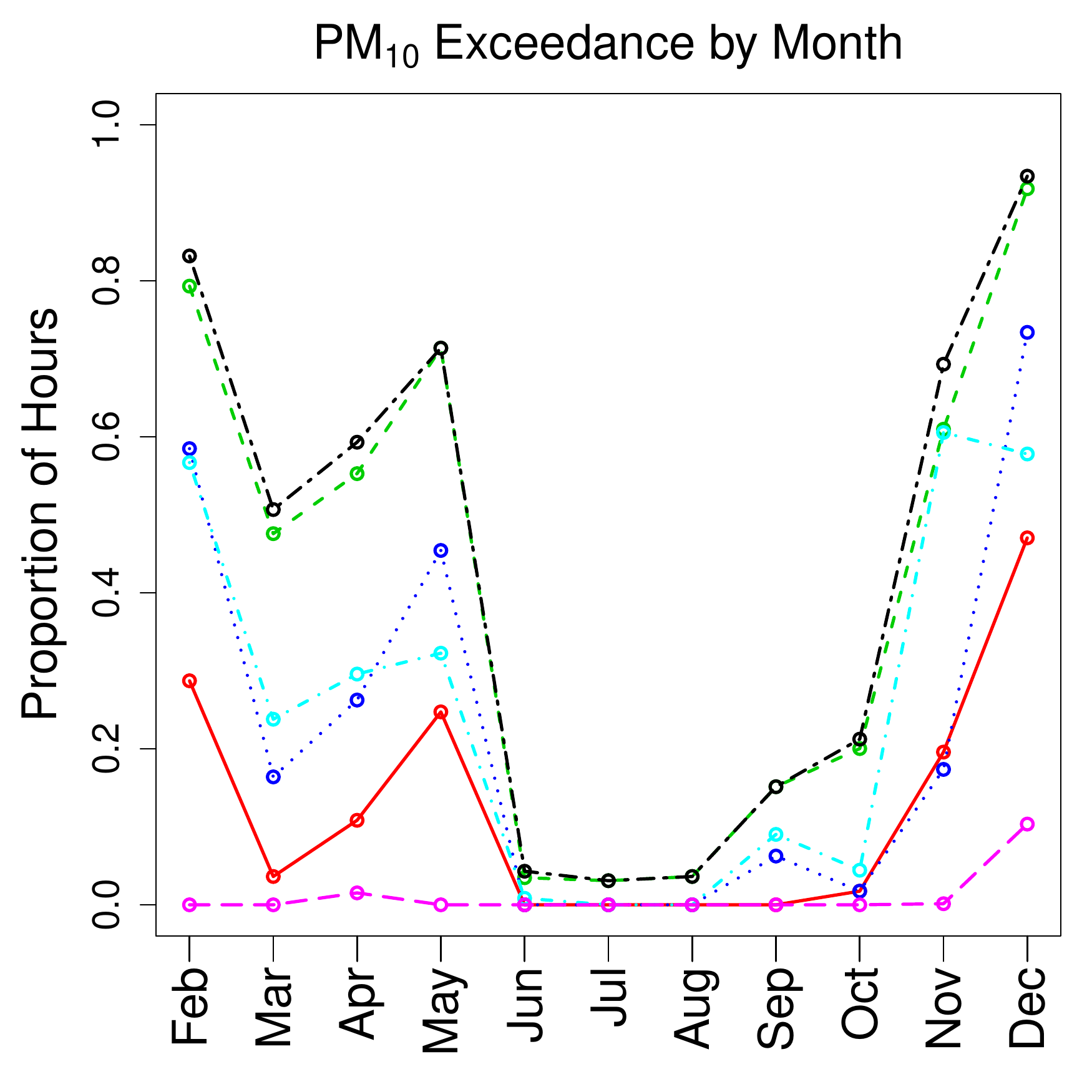}
      \subcaption{$\text{PM}_{10}$ exceedance by month}\label{fig:legislated_pm}
   \end{subfigure}
     \begin{subfigure}[b]{.32\textwidth}
     \includegraphics[width=\textwidth]{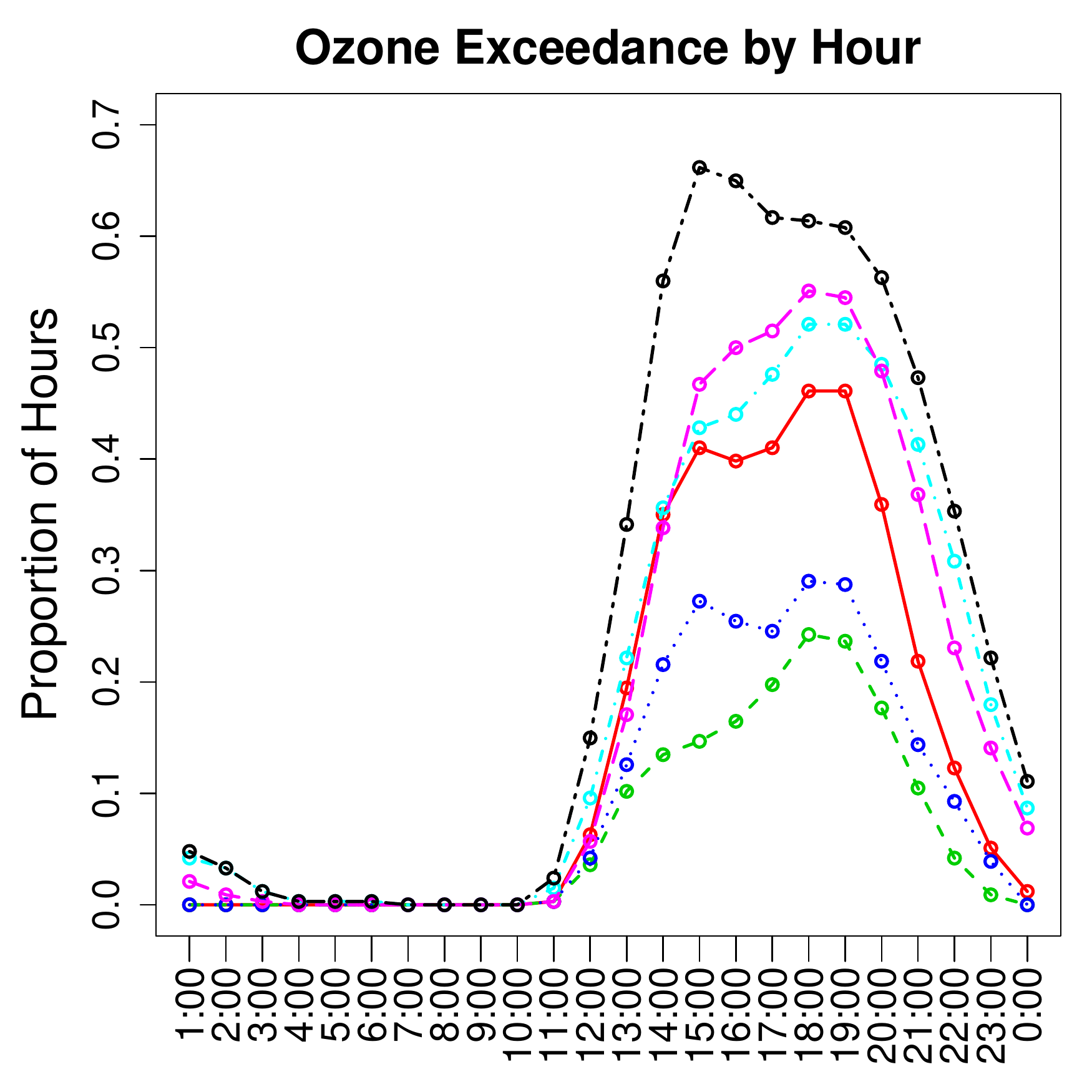}
      \subcaption{Time of day $\text{O}_{3}$ exceeded}\label{fig:legislated_ozone_hour}
   \end{subfigure}
  \end{center}
  \vspace{-4mm}
       \caption{Posterior predictive means for the proportion of hours of exceedance as function of month and hour of the day. }\label{fig:legislated}
\end{figure}

We continue the prospective analysis for all months except January, fitting the model up until the last hour of the previous month to predict pollution exceedance for the month of interest. Because we fit the model sequentially, prospective predictions for January are poor because the model has not been trained on data for these times.
%\textbf{I don't know what ``informative'' means here?  And, according to our prediction strategy, I would start with February, using  model fitting results, from data up to the last hour of January} 
Using these predictions, we give posterior means and 95\% credible intervals for the one-hour-ahead predicted proportion of hours and days of exceedance for each region (i.e. $P(Z_{jt}^O > 95 \text{ ppb} \,\cup \, Z_{jt}^{\overline{O}} > 70 \text{ ppb} )$, $P(Z_{jt}^{PM} >  75 \text{ } \mu g/m^3)$, $ P(W_{jd}^{O}  > 95 \text{ ppb}  \cup W_{jd}^{\overline{O}} > 70 \text{ ppb} )$, and $P(W_{jd}^{PM} >  75 \text{ } \mu g/m^3)$, as defined in Section \ref{sec:post_analysis}). The results for ozone are given in Table \ref{tab:ozone_region}, and the estimates for $\text{PM}_{10}$ are presented in Table \ref{tab:pm10_region}. For ozone, the proportion of exceedances in both hours and days decreases as latitude increases, with northern regions showing nearly half as many exceedances as the southern regions, on average. The trend for $\text{PM}_{10}$ is less clear, although there is significant variability across regions. The northeast region has many more predicted $\text{PM}_{10}$ exceedances than any other region. This is due to the large industrial economy located within this region. By contrast, the southwest region has, comparatively, very few $\text{PM}_{10}$ exceedances.
%\textbf{Do you think proportions of total rather than hours would look better of in Tables 6 and 7?}
\vspace{-3mm}
\begin{table}[H]
\centering
\footnotesize
\begin{tabular}{l|rrrrrr|rrrrrr}
\hline
Ozone &\multicolumn{6}{c|}{Hours (total of 8016) } & \multicolumn{6}{|c}{Days (total of 334)}\\
 & CE & NE & NW & SE & SW & Any & CE & NE & NW & SE & SW & Any \\
  \hline
% & 1175 & 532 & 745 & 1552 & 1492 & 2022 & 217 & 123 & 167 & 224 & 232 & 265 \\ 
%  2.5\% & 1149 & 511 & 721 & 1526 & 1466 & 1997 & 209 & 113 & 156 & 216 & 225 & 258 \\ 
%  97.5\% & 1202 & 553 & 770 & 1579 & 1518 & 2048 & 226 & 132 & 177 & 233 & 240 & 272 \\ 
Mean & 0.147 & 0.066 & 0.093 & 0.194 & 0.186 & 0.252 & 0.651 & 0.367 & 0.499 & 0.671 & 0.696 & 0.794 \\ 
  2.5\% & 0.143 & 0.064 & 0.090 & 0.190 & 0.183 & 0.249 & 0.626 & 0.338 & 0.467 & 0.647 & 0.674 & 0.773 \\ 
  97.5\% & 0.150 & 0.069 & 0.096 & 0.197 & 0.189 & 0.256 & 0.677 & 0.395 & 0.530 & 0.698 & 0.719 & 0.814 \\ 
   \hline
\end{tabular}
\caption{One-hour-ahead posterior predictive estimates for the (Left) Proportion of hours where either of the Mexican legislated ozone limits (one-hour or eight-hour) were exceeded (Right) Proportion of days where either of the Mexican legislated ozone limits were exceeded. Posterior means and 95\% credible intervals are given for each region. The ``Any'' label indicates that at least one region has an exceedance for one or more location for the time level (hour or day). }\label{tab:ozone_region}
\end{table}
\vspace{-3mm}
\begin{table}[H]
\centering
\footnotesize
\begin{tabular}{l|rrrrrr|rrrrrr}
\hline
$\text{PM}_{10}$ &\multicolumn{6}{c|}{Hours (total of 8016) } & \multicolumn{6}{|c}{Days (total of 334)}\\
 & CE & NE & NW & SE & SW & Any & CE & NE & NW & SE & SW & Any \\
  \hline
% & 987 & 3271 & 1771 & 1980 & 90 & 3436 & 73 & 174 & 111 & 123 & 9 & 179 \\ 
%  2.5\% & 974 & 3257 & 1757 & 1965 & 85 & 3422 & 70 & 171 & 108 & 121 & 7 & 175 \\ 
%  97.5\% & 1001 & 3284 & 1786 & 1995 & 95 & 3450 & 76 & 178 & 114 & 126 & 10 & 182 \\ 
 Mean & 0.123 & 0.408 & 0.221 & 0.247 & 0.0112 & 0.429 & 0.218 & 0.5223 & 0.333 & 0.370 & 0.026 & 0.535 \\ 
  2.5\% & 0.122 & 0.406 & 0.219 & 0.245 & 0.0106 & 0.427 & 0.210 & 0.512 & 0.323 & 0.362 & 0.021 & 0.524 \\ 
  97.5\% & 0.125 & 0.410 & 0.223 & 0.249 & 0.0119 & 0.430 & 0.228 & 0.533 & 0.341 & 0.377 & 0.030 & 0.545 \\ 
   \hline
\end{tabular}
\caption{One-hour-ahead posterior predictive estimates for the (Left) Proportion of hours where the Mexican legislated 24-hour $\text{PM}_{10}$ limits were exceeded (Right) Proportion of days where either of the Mexican legislated $\text{PM}_{10}$ limits were exceeded. The ``Any'' label indicates that at least one region has an exceedance for one or more location for the time level (hour or day). }\label{tab:pm10_region}
\end{table}
\vspace{-3mm}
Lastly, we discuss the proportion of predicted hourly exceedances as a function of the month of the year and of the hour of the day. The summaries for ozone by month and hour-of-day are plotted in Figure \ref{fig:legislated}, while we display $\text{PM}_{10}$ exceedances only by month (Figure \ref{fig:legislated_pm}). We do not plot $\text{PM}_{10}$ exceedances as function of the hour of the day because $\text{PM}_{10}$ exceedances depend on 24-hour averages; thus, trends over time-of-day are not meaningful. As a function of month, the patterns of ozone and $\text{PM}_{10}$ exceedances are clear. For ozone, the proportion of exceedances reaches a peak in May and is high in March, April, and June. We attribute these high ozone levels to warm times of the year that are dry compared to the rainy season (June-August).  $\text{PM}_{10}$ exceedance appears to co-vary strongly with the rainy season as well, which is captured by relative humidity in our model. In particular, June, July, August, and September have almost no exceedances for $\text{PM}_{10}$. The coldest months (December and February) have higher probabilities of $\text{PM}_{10}$ exceedance than warmer months that are similarly dry. Mexico City's pollution output is higher during winter festivities like Our Lady of Guadalupe, Christmas, and New Year due to fireworks and increased motor traffic. In conjunction with increased pollution output, pollution exceedances in cold months are also due to thermal inversion that traps pollution in the Valley of Mexico where Mexico City lies. Ozone exceedances also tend to peak in the afternoon to evening. Because ozone levels can exceed thresholds for either one-hour or eight-hour average ozone, we expect two peaks in ozone exceedance as a function of hour. The one-hour peak occurs around 4 PM (16:00) when the temperature is highest. The peak of eight-hour average ozone peaks around 7 or 8 PM (19:00 or 20:00), after eight hours of relatively high ozone levels. These peaks can be seen in Figure \ref{fig:legislated_ozone_hour}.
\vspace{-5mm}

\section{Conclusions and Future Work}\label{sec:conc}

We have discussed the monitoring network for ozone and $\text{PM}_{10}$ within Mexico City and proposed a joint spatiotemporal model for ozone and $\text{PM}_{10}$ concentrations. This model was used to predict future pollutant concentrations. Our predictions were then used to obtain derived distributions for regional maxima of ozone and $\text{PM}_{10}$ which are needed to determine Mexico City's pollution emergency phases. Additionally, our predictions are used to assess compliance with MAAQS. We find that predicted risk of pollution emergency is rare and are predicted for only a few periods of 2017. By contrast, we demonstrate that predicted exceedance of Mexico's ambient air quality standards is common.

In future work, we will attempt to operationalize our model so that it can be used in practice. This would require real time (hourly) fitting of the model as new measurements are available. Our modeling is amenable to sequential updating as well as possible parallelization though considerable optimization remains before this could be implemented in practice. Once implemented, our model could warn of potential pollution emergencies or compliance issues, allowing regional and city-wide adjustments, warnings, responses, and decision-making to be made earlier. Our model could incorporate weather forecasts to perhaps more accurately forecast pollutant levels farther ahead than our one-hour-ahead predictions.

%\section*{Acknowledgments}

%\bibliographystyle{rss}
%\bibliographystyle{environmetrics}
\bibliography{refs}

\appendix

\section{Full Conditional Distributions for AR Model}\label{app:gibbs3}

We give the full conditional distributions for the model specified in Section \ref{sec:model}. We give some additional details here to clarify model fitting. Because $\bV_1$ and $\bV_2$ are independent \emph{a priori}, the joint prior distribution for $\bV_1$ and $\bV_2$ is
\begin{equation}
[\bV_1,\bV_2 ] \propto \exp\left(- \frac{1}{2 } \bV_1^T Q \bV_1  \right) \exp\left(- \frac{1}{2 } \bV_2^T Q \bV_2 \right).
\end{equation}
The induced joint prior distribution of $\begin{pmatrix}
\bpsi_{1} \\ \bpsi_{2} \end{pmatrix}$, where $\bpsi_1 = (\psi_{11},\psi_{12},...,\psi_{1N_s} )^T$ and $\bpsi_2 = (\psi_{21},\psi_{22},...,\psi_{2N_s} )^T$, is used for model fitting and can be represented as
\begin{align}
&[\bpsi_1,\bpsi_2 |A_\psi] = [\bpsi_1 | A_\psi] [\bpsi_2 | \bpsi_1,A_\psi] \nonumber \\
&\propto \exp\left(- \frac{1}{2 {a_{11}^{(\psi)}}^2 }  \bpsi_1^T Q \bpsi_1  \right) \exp\left(- \frac{1}{2 {a_{22}^{(\psi)}}^2 } \left(\bpsi_2 - \frac{a_{12}^{(\psi)}}{a_{11}^{(\psi)}} \bpsi_1 \right)^T Q \left(\bpsi_2 - \frac{a_{12}^{(\psi)}}{a_{11}^{(\psi)}} \bpsi_1 \right) \right) .
\end{align}

For this section, let $\theta | \cdots$ indicate the full conditional distribution of $\theta$, where $\theta$ is an arbitrary parameter. For several quantities, we combine site-specific variables. For example, let $Y_t^O = (Y_{1t}^O,...,Y_{N_s t}^O)^T$, $Y_t^{PM} = (Y_{1t}^{PM},...,Y_{N_s t}^{PM})^T$, $\bX_t = \text{blockdiag}(\bx_{it} )$ and $\bbeta_k = (\bbeta_{k1} ,...,\bbeta_{k N_s})^T$. In addition to previous terms, we also let $\bL_t = \text{blockdiag}(\bL_{it})$ and $\bgamma_k = (\bgamma_{k1} ,...,\bgamma_{k N_s})^T$. The full conditional distributions for this model are provided below.
\begin{equation*}
\begin{aligned}[l]
\beta_{1i} | \cdots &\sim N(V_{\beta_{1i} }^* m_{\beta_{1i} }^* , V_{\beta_{1i} }^*) \\
\beta_{2i} | \cdots &\sim N(V_{\beta_{2i} }^* m_{\beta_{2i} }^* , V_{\beta_{2i} }^*) \\
\beta_{01} | \cdots &\sim N(V_{\beta_{01} }^* m_{\beta_{01} }^* , V_{\beta_{01} }^*) \\
\beta_{02} | \cdots &\sim N(V_{\beta_{02} }^* m_{\beta_{02} }^* , V_{\beta_{02} }^*) \\
\Sigma_{\beta_{1}} | \cdots &\sim IW(M_{\beta_{1}}^*,\nu^*_{\beta_{1}}) \\
\Sigma_{\beta_{2}}  | \cdots &\sim IW(M_{\beta_{2}}^*,\nu_{\beta_{2}}^*) 
\end{aligned}
\hspace{6mm}
\begin{aligned}[l]
\gamma_{1i} | \cdots &\sim N(V_{\gamma_{1i} }^* m_{\gamma_{1i} }^* , V_{\gamma_{1i} }^*) \\
\gamma_{2i} | \cdots &\sim N(V_{\gamma_{2i} }^* m_{\gamma_{2i} }^* , V_{\gamma_{2i} }^*) \\
\gamma_{01} | \cdots &\sim N(V_{\gamma_{01} }^* m_{\gamma_{01} }^* , V_{\gamma_{01} }^*) \\
\gamma_{02} | \cdots &\sim N(V_{\gamma_{02} }^* m_{\gamma_{02} }^* , V_{\gamma_{02} }^*) \\
\Sigma_{\gamma_{1}} | \cdots &\sim IW(M_{\gamma_1}^*,\nu^*_{\gamma_1}) \\
\Sigma_{\gamma_{2}}  | \cdots &\sim IW(M_{\gamma_2}^*,\nu_{\gamma_2}^*) 
\end{aligned}
\hspace{6mm}
\begin{aligned}[l]
\sigma^2_1 | \cdots &\sim IG(a^*_{\sigma_1},b^*_{\sigma_1}) \\
\sigma^2_2 | \cdots &\sim IG(a^*_{\sigma_2},b^*_{\sigma_2}) \\
\bV_1 | \cdots &\sim N(V_{V_1}^* m_{V_1}^* , V_{V_1}^*) \\
\bV_2 | \cdots &\sim N(V_{V_2}^* m_{V_2}^* , V_{V_2}^*) \\
{a^{(\psi)}_{11}}^2  | \cdots &\sim IG(a_{a_1}^*,b_{a_1}^*), \\
{a^{(\psi)}_{22}}^2 | \cdots   &\sim IG(a_{a_2}^*,b_{a_2}^*), \\
a^{(\psi)}_{12} | \cdots &\sim N(V_{\psi}^* m_{\psi}^* , V_{\psi}^*) 
\end{aligned}
\end{equation*}
with
{\footnotesize
\begin{align*}
a^*_{\sigma_1} &= 1 + \frac{N_s \times N_t}{2}\\
b^*_{\sigma_1} &= 1 + \frac{1}{2} \sum_{t=1}^{N_t} \sum_{i=1}^{N_s} (Y^{O}_{it} - \bx_{it}^T \bbeta_{1i}-{\bL^{O}_{it}}^T\bgamma_{1i} - a_{11}^{(\psi)} V_{1i}  )^2 \\
a^*_{\sigma_2} &= 1 + \frac{N_s \times N_t}{2}\\
b^*_{\sigma_2} &= 1 + \frac{1}{2} \sum_{t=1}^{N_t} \sum_{i=1}^{N_s} (Y^{PM}_{it} - \bx_{it}^T \bbeta_{2i} -{\bL^{PM}_{it}}^T\bgamma_{2i} - a_{12}^{(\psi)} V_{1i}- a_{22}^{(\psi)} V_{2i}  )^2 \\
m_{\beta_{1i} }^* &= \Sigma_{\beta_1}^{-1} \beta_{01} + \frac{1}{\sigma^2_1} \sum^{N_t}_{t=1} \bx_{it} (Y^{O}_{it}- {\bL^{O}_{it}}^T\bgamma_{1i} - a_{11}^{(\psi)} V_{1i} )   \\
V_{\beta_{1i} }^* &= \left( \Sigma_{\beta_1}^{-1} + \frac{1}{\sigma_1^2} \sum^{N_t}_{t=1} \bx_{it} \bx_{it}^T  \right)^{-1} \\
m_{\beta_{2i} }^* &= \Sigma_{\beta_2}^{-1} \beta_{02} + \frac{1}{\sigma^2_2} \sum^{N_t}_{t=1} \bx_{it} (Y^{PM}_{it} - {\bL^{PM}_{it}}^T\bgamma_{2i} - a_{12}^{(\psi)} V_{1i}- a_{22}^{(\psi)} V_{2i}  )   \\ \\
V_{\beta_{2i} }^* &= \left( \Sigma_{\beta_2}^{-1} + \frac{1}{\sigma_2^2} \sum^{N_t}_{t=1} \bx_{it} \bx_{it}^T  \right)^{-1} \\
m_{\beta_{01} }^* &=  \sum_{i=1}^{N_s}  \Sigma_{\beta_1}^{-1} \bbeta_{1i} \\
V_{\beta_{01} }^* &= \left( N_s \Sigma_{\beta_1}^{-1}  + 10^{-3} \, \bI \right)^{-1} \\
m_{\beta_{02} }^* &=  \sum_{i=1}^{N_s}  \Sigma_{\beta_2}^{-1} \bbeta_{2i}  \\
V_{\beta_{02} }^* &= \left( N_s \Sigma_{\beta_2}^{-1}  + 10^{-3} \, \bI \right)^{-1}  \\
M_{\beta_1}^* &= 10^{3} \, \bI + \sum_{i=1}^{N_s} (\bbeta_{1i} - \bbeta_{01}) (\bbeta_{1i} - \bbeta_{01})^T \\
\nu_{\beta_1}^* &= N_s + p + 1 \\
M_{\beta_2}^* &=  10^{3} \, \bI + \sum_{i=1}^{N_s} (\bbeta_{2i} - \bbeta_{02}) (\bbeta_{2i} - \bbeta_{02})^T \\
\nu_{\beta_2}^* &= N_s + p + 1 \\
m_{\gamma_{1i} }^* &= \Sigma_{\gamma_1}^{-1} \bgamma_{01} + \frac{1}{\sigma^2_1} \sum^{N_t}_{t=1} {\bL^{O}_{it}} (Y^{O}_{it}- \bx_{it}^T\bbeta_{1i} - a_{11}^{(\psi)} V_{1i} )   \\
V_{\gamma_{1i} }^* &= \left( \Sigma_{\gamma_1}^{-1} + \frac{1}{\sigma_1^2} \sum^{N_t}_{t=1} {\bL^{O}_{it}} {\bL^{O}_{it}}^T  \right)^{-1} \\
m_{\gamma_{2i} }^* &= \Sigma_{\gamma_2}^{-1} \bgamma_{02} + \frac{1}{\sigma^2_2} \sum^{N_t}_{t=1} {\bL^{PM}_{it}} (Y^{PM}_{it} - \bx_{it}^T\bbeta_{2i} - a_{12}^{(\psi)} V_{1i}- a_{22}^{(\psi)} V_{2i} )   \\
\Sigma_{\gamma_{2i} }^* &= \left( \Sigma_{\gamma_2}^{-1} + \frac{1}{\sigma_2^2} \sum^{N_t}_{t=1} {\bL^{PM}_{it}} {\bL^{PM}_{it}}^T  \right)^{-1} \\
m_{\gamma_{01} }^* &=  \sum_{i=1}^{N_s}  \Sigma_{\gamma_1}^{-1} \bgamma_{1i} \\
V_{\gamma_{01} }^* &= \left( N_s \Sigma_{\gamma_1}^{-1}  +10^{-3} \, \bI \right)^{-1} \\
m_{\gamma_{02} }^* &=  \sum_{i=1}^{N_s}  \Sigma_{\gamma_2}^{-1} \bgamma_{2i}  \\
V_{\gamma_{02} }^* &= \left( N_s \Sigma_{\gamma_2}^{-1}  + 10^{-3} \, \bI \right)^{-1}  \\
M_{\gamma_1}^* &=10^{3} \, \bI + \sum_{i=1}^{N_s} (\bgamma_{1i} - \bgamma_{01}) (\bgamma_{1i} - \bgamma_{01})^T \\
\nu_{\gamma_1}^* &= N_s + n_{1l} + 1 \\
M_{\gamma_2}^* &= 10^{3} \, \bI + \sum_{i=1}^{N_s} (\bgamma_{2i} - \bgamma_{02}) (\bgamma_{2i} - \bgamma_{02})^T \\
\nu_{\gamma_2}^* &= N_s + n_{2l} + 1 \\
m_{V_1}^* &= \frac{a_{11}^{(\psi)} }{\sigma^2_1} \sum_{i=1}^{N_t} (Y_t^{O} - \bX_t \bbeta_1- \bL^{O}_t\bgamma_{1} ) + \\ & \frac{a_{12}^{(\psi)} }{\sigma^2_2} \sum_{i=1}^{N_t} (Y_{t}^{PM} - \bX_t \bbeta_2- \bL^{PM}_t\bgamma_{2}- a_{22}^{(\psi)} \bV_2  )  \\
V_{V_1}^* &= \left( Q_1 + \left[\frac{{a_{11}^{(\psi)}}^2 N_t }{\sigma^2_1} + \frac{{a_{12}^{(\psi)}}^2 N_t }{\sigma^2_2} \right] \bI \right)^{-1} \\
m_{V_2}^* &= \frac{a_{22}^{(\psi)} }{\sigma^2_2} \sum_{i=1}^{N_t} (Y_t^{PM} - \bX_t \bbeta_2-\bL^{PM}_t\bgamma_{2}- a_{12}^{(\psi)} \bV_1 ) \\
V_{V_2}^* &= \left( Q_2  + \bI \frac{{a_{22}^{(\psi)}}^2 N_t }{\sigma^2_2} \right)^{-1} \\
a^*_{a_1} &= 1 + N_s /2  \\
b^*_{a_1} &= 1 + \frac{1}{2} \bpsi_1^T Q_1 \bpsi_1 \\
a^*_{a_2} &= 1 + N_s /2   \\
b^*_{a_2} &= 1+ \frac{1}{2} \left(\bpsi_2 - \frac{a_{12}^{(\psi)}}{a_{11}^{(\psi)}} \bpsi_1 \right)^T Q \left(\bpsi_2 - \frac{a_{12}^{(\psi)}}{a_{11}^{(\psi)}} \bpsi_1 \right) \\
 m_{\psi}^* &= \frac{1}{\sigma^2_2} \sum^{N_t}_{t=1} \sum^{N_s}_{i=1} V_{1i} (Y^{PM}_{it} - {\bL^{PM}_{it}}^T\bgamma_{2i}- \bx_{it}^T \beta_{2i} - a^{(\psi)}_{22} V_{2i}  )\\
V_{\psi}^* &= \left( 10^{-3} + \frac{1}{\sigma^2_2} \sum^{N_t}_{t=1} \sum^{N_s}_{i=1} V_{1i}^2 \right)^{-1} 
\end{align*}
}

\section{Full Conditional Distributions for Heteroscedastic AR Model}\label{app:gibbs4}

We rely on some of the details presented in Appendix \ref{app:gibbs3} for the CAR terms. We give the full conditional distributions for the heteroscedastic model specified in Section \ref{sec:model} where the variance is a function of the hour of the day $h(t)$. For this section, let $\theta | \cdots$ indicate the full conditional distribution of $\theta$, where $\theta$ is an arbitrary parameter. We again combine several quantities for site-specific variables. Let $Y_t^O = (Y_{1t}^O,...,Y_{N_s t}^O)^T$, $Y_t^{PM} = (Y_{1t}^{PM},...,Y_{N_s t}^{PM})^T$, $\bX_t = \text{blockdiag}(\bx_{it} )$ and $\bbeta_k = (\bbeta_{k1} ,...,\bbeta_{k N_s})^T$. In addition to previous terms, we also let $\bL_t = \text{blockdiag}(\bL_{it})$ and $\bgamma_k = (\bgamma_{k1} ,...,\bgamma_{k N_s})^T$. The full conditional distributions for this model are provided below. If posterior parameters are not given below, then they are identical to those given for the homoscedastic model in Appendix \ref{app:gibbs3}.
\begin{equation*}
\begin{aligned}[l]
\beta_{1i} | \cdots &\sim N(V_{\beta_{1i} }^* m_{\beta_{1i} }^* , V_{\beta_{1i} }^*) \\
\beta_{2i} | \cdots &\sim N(V_{\beta_{2i} }^* m_{\beta_{2i} }^* , V_{\beta_{2i} }^*) \\
\beta_{01} | \cdots &\sim N(V_{\beta_{01} }^* m_{\beta_{01} }^* , V_{\beta_{01} }^*) \\
\beta_{02} | \cdots &\sim N(V_{\beta_{02} }^* m_{\beta_{02} }^* , V_{\beta_{02} }^*) \\
\Sigma_{\beta_{1}} | \cdots &\sim IW(M_{\beta_{1}}^*,\nu^*_{\beta_{1}}) \\
\Sigma_{\beta_{2}}  | \cdots &\sim IW(M_{\beta_{2}}^*,\nu_{\beta_{2}}^*) 
\end{aligned}
\hspace{6mm}
\begin{aligned}[l]
\gamma_{1i} | \cdots &\sim N(V_{\gamma_{1i} }^* m_{\gamma_{1i} }^* , V_{\gamma_{1i} }^*) \\
\gamma_{2i} | \cdots &\sim N(V_{\gamma_{2i} }^* m_{\gamma_{2i} }^* , V_{\gamma_{2i} }^*) \\
\gamma_{01} | \cdots &\sim N(V_{\gamma_{01} }^* m_{\gamma_{01} }^* , V_{\gamma_{01} }^*) \\
\gamma_{02} | \cdots &\sim N(V_{\gamma_{02} }^* m_{\gamma_{02} }^* , V_{\gamma_{02} }^*) \\
\Sigma_{\gamma_{1}} | \cdots &\sim IW(M_{\gamma_1}^*,\nu^*_{\gamma_1}) \\
\Sigma_{\gamma_{2}}  | \cdots &\sim IW(M_{\gamma_2}^*,\nu_{\gamma_2}^*) 
\end{aligned}
\hspace{6mm}
\begin{aligned}[l]
\sigma^2_{1 q} | \cdots &\sim IG(a^*_{\sigma_{1 q}},b^*_{\sigma_{1 q} }) \\
\sigma^2_{2q} | \cdots &\sim IG(a^*_{\sigma_{2q}},b^*_{\sigma_{2q}}) \\
\bV_1 | \cdots &\sim N(V_{V_1}^* m_{V_1}^* , V_{V_1}^*) \\
\bV_2 | \cdots &\sim N(V_{V_2}^* m_{V_2}^* , V_{V_2}^*) \\
{a^{(\psi)}_{11}}^2  | \cdots &\sim IG(a_{a_1}^*,b_{a_1}^*), \\
{a^{(\psi)}_{22}}^2 | \cdots   &\sim IG(a_{a_2}^*,b_{a_2}^*), \\
a^{(\psi)}_{12} | \cdots &\sim N(V_{\psi}^* m_{\psi}^* , V_{\psi}^*)  \\
\end{aligned}
\end{equation*}
with

{\footnotesize
\begin{align*}
a^*_{\sigma_{1 q}} &= 1 + \frac{N_s \times N_t}{2 N_h }\\
b^*_{\sigma_{1 q}} &= 1 + \frac{1}{2} \sum_{t=1}^{N_t} \sum_{i=1}^{N_s} (Y^{O}_{it} - \bx_{it}^T \bbeta_{1i}-{\bL^{O}_{it}}^T\bgamma_{1i} - a_{11}^{(\psi)} V_{1i}  )^2 \bone( h(t) = q) \\
a^*_{\sigma_{2 q}} &= 1 + \frac{N_s \times N_t}{2 N_h}\\
b^*_{\sigma_{2 q}} &= 1 + \frac{1}{2} \sum_{t=1}^{N_t} \sum_{i=1}^{N_s} (Y^{PM}_{it} - \bx_{it}^T \bbeta_{2i} -{\bL^{PM}_{it}}^T\bgamma_{2i} - a_{12}^{(\psi)} V_{1i}- a_{22}^{(\psi)} V_{2i}  )^2\bone( h(t) = q) \\
m_{\beta_{1i} }^* &= \Sigma_{\beta_1}^{-1} \beta_{01} + \sum^{N_t}_{t=1} \frac{1}{\sigma^2_{1h(t)} }  \bx_{it} (Y^{O}_{it}- {\bL^{O}_{it}}^T\bgamma_{1i} - a_{11}^{(\psi)} V_{1i} )   \\
V_{\beta_{1i} }^* &= \left( \Sigma_{\beta_1}^{-1} + \sum^{N_t}_{t=1}\frac{1}{\sigma^2_{1h(t)} }  \bx_{it} \bx_{it}^T  \right)^{-1} \\
m_{\beta_{2i} }^* &= \Sigma_{\beta_2}^{-1} \beta_{02} +\sum^{N_t}_{t=1} \frac{1}{\sigma^2_{2h(t)} } \bx_{it} (Y^{PM}_{it} - {\bL^{PM}_{it}}^T\bgamma_{2i} - a_{12}^{(\psi)} V_{1i}- a_{22}^{(\psi)} V_{2i}  )   \\ \\
V_{\beta_{2i} }^* &= \left( \Sigma_{\beta_2}^{-1} +\sum^{N_t}_{t=1}  \frac{1}{\sigma^2_{2h(t)} }\bx_{it} \bx_{it}^T  \right)^{-1} \\
%m_{\beta_{01} }^* &= \bS_{\beta_1}^{-1} \bm_{\beta_1} +  \sum_{i=1}^{N_s}  \Sigma_{\beta_1}^{-1} \bbeta_{1i} \\
%V_{\beta_{01} }^* &= \left( N_s \Sigma_{\beta_1}^{-1}  + \bS_{\beta_1}^{-1} \right)^{-1} \\
%m_{\beta_{02} }^* &= \bS_{\beta_2}^{-1} \bm_{\beta_2} +  \sum_{i=1}^{N_s}  \Sigma_{\beta_2}^{-1} \bbeta_{2i}  \\
%V_{\beta_{02} }^* &= \left( N_s \Sigma_{\beta_2}^{-1}  + \bS_{\beta_2}^{-1} \right)^{-1}  \\
%M_{\beta_1}^* &= M_{\beta_1} + \sum_{i=1}^{N_s} (\bbeta_{1i} - \bbeta_{01}) (\bbeta_{1i} - \bbeta_{01})^T \\
%\nu_{\beta_1}^* &= N_s + \nu_{\beta_1} \\
%M_{\beta_2}^* &=  M_{\beta_2} + \sum_{i=1}^{N_s} (\bbeta_{2i} - \bbeta_{02}) (\bbeta_{2i} - \bbeta_{02})^T \\
%\nu_{\beta_2}^* &= N_s + \nu_{\beta_2} \\
m_{\gamma_{1i} }^* &= \Sigma_{\gamma_1}^{-1} \bgamma_{01} + \sum^{N_t}_{t=1}  \frac{1}{\sigma^2_{1h(t)} } {\bL^{O}_{it}} (Y^{O}_{it}- \bx_{it}^T\bbeta_{1i} - a_{11}^{(\psi)} V_{1i} )   \\
V_{\gamma_{1i} }^* &= \left( \Sigma_{\gamma_1}^{-1} + \sum^{N_t}_{t=1} \frac{1}{\sigma^2_{1h(t)} } {\bL^{O}_{it}} {\bL^{O}_{it}}^T  \right)^{-1} \\
m_{\gamma_{2i} }^* &= \Sigma_{\gamma_2}^{-1} \bgamma_{02} + \sum^{N_t}_{t=1}  \frac{1}{\sigma^2_{2h(t)} }{\bL^{PM}_{it}} (Y^{PM}_{it} - \bx_{it}^T\bbeta_{2i} - a_{12}^{(\psi)} V_{1i}- a_{22}^{(\psi)} V_{2i} )   \\
V_{\gamma_{2i} }^* &= \left( \Sigma_{\gamma_2}^{-1} + \sum^{N_t}_{t=1} \frac{1}{\sigma^2_{2h(t)} } {\bL^{PM}_{it}} {\bL^{PM}_{it}}^T  \right)^{-1} \\
%m_{\gamma_{01} }^* &= \bS_{\gamma_1}^{-1} \bm_{\gamma_1} +  \sum_{i=1}^{N_s}  \Sigma_{\gamma_1}^{-1} \bgamma_{1i} \\
%V_{\gamma_{01} }^* &= \left( N_s \Sigma_{\gamma_1}^{-1}  + \bS_{\gamma_1}^{-1} \right)^{-1} \\
%m_{\gamma_{02} }^* &= \bS_{\gamma_2}^{-1} \bm_{\gamma_2} +  \sum_{i=1}^{N_s}  \Sigma_{\gamma_2}^{-1} \bgamma_{2i}  \\
%V_{\gamma_{02} }^* &= \left( N_s \Sigma_{\gamma_2}^{-1}  + \bS_{\gamma_2}^{-1} \right)^{-1}  \\
%M_{\gamma_1}^* &= M_{\gamma_1} + \sum_{i=1}^{N_s} (\bgamma_{1i} - \bgamma_{01}) (\bgamma_{1i} - \bgamma_{01})^T \\
%\nu_{\gamma_1}^* &= N_s + \nu_{\gamma_1} \\
%M_{\gamma_2}^* &=  M_{\gamma_2} + \sum_{i=1}^{N_s} (\bgamma_{2i} - \bgamma_{02}) (\bgamma_{2i} - \bgamma_{02})^T \\
%\nu_{\gamma_2}^* &= N_s + \nu_{\gamma_2} \\
m_{V_1}^* &=  a_{11}^{(\psi)}  \sum_{t=1}^{N_t}   \frac{1}{\sigma^2_{1h(t)} }(Y_t^{O} - \bX_t \bbeta_1- \bL^{O}_t\bgamma_{1} ) + \\ & a_{12}^{(\psi)} \sum_{t=1}^{N_t} \frac{1}{\sigma^2_{2h(t)} } (Y_t^{PM} - \bX_t \bbeta_2- \bL^{PM}_t\bgamma_{2}- a_{22}^{(\psi)} \bV_2  )  \\
V_{V_1}^* &= \left( Q_1 + \left[{a_{11}^{(\psi)}}^2 \frac{N_t}{N_h} \sum_{q=1}^{24} \sigma^2_{1q} + {a_{12}^{(\psi)}}^2 \frac{ N_t }{N_h} \sum_{q=1}^{24} \sigma^2_{2q} \right] \bI \right)^{-1} \\
m_{V_2}^* &= \frac{a_{22}^{(\psi)} }{\sigma^2_2} \sum_{i=1}^{N_t} (Y_t^{PM} - \bX_t \bbeta_2-\bL^{PM}_t\bgamma_{2}- a_{12}^{(\psi)} \bV_1 ) \\
V_{V_2}^* &= \left( Q_2 + \bI {a_{22}^{(\psi)}}^2 \frac{ N_t }{N_h} \sum_{q=1}^{24} \sigma^2_{2q} \right)^{-1} \\
%a^*_{a_1} &=  N_s /2  + a_{a_1} \\
%b^*_{a_1} &= b_{a_1} + \frac{1}{2} \bpsi_1^T Q_1 \bpsi_1 \\
%a^*_{a_2} &=  N_s /2  + a_{a_2} \\
%b^*_{a_2} &= b_{a_2} + \frac{1}{2} \left(\bpsi_2 - \frac{a_{12}^{(\psi)}}{a_{11}^{(\psi)}} \bpsi_1 \right)^T Q \left(\bpsi_2 - \frac{a_{12}^{(\psi)}}{a_{11}^{(\psi)}} \bpsi_1 \right) \\
 m_{\psi}^* &= \sum^{N_t}_{t=1} \frac{1}{\sigma^2_{2h(t)} } \sum^{N_s}_{i=1}  V_{1i} (Y^{PM}_{it} - {\bL^{PM}_{it}}^T\bgamma_{2i}- \bx_{it}^T \beta_{2i} - a^{(\psi)}_{22} V_{2i}  )\\
V_{\psi}^* &= \left( 10^{-3} +  \sum^{N_t}_{t=1} \frac{1}{\sigma^2_{2h(t)} }  \sum^{N_s}_{i=1} V_{1i}^2 \right)^{-1} 
\end{align*}
}
\section{Prediction of Held-out Data}\label{app:full_cond_pred}

For model validation, we hold out 10\% of the data and impute or update these held-out values each step of the Gibbs sampler which is described below.
\begin{align*}
\mu_{1it} &= \bx_{i(t-1)}^T \bbeta_{1i} + {\bL^{O}_{it}}^T\bgamma_{1i}+ \psi_{1i}, \\
\mu_{2it} &= \bx_{i(t-1)}^T \bbeta_{2i} + {\bL^{PM}_{it}}^T\bgamma_{2i}+ \psi_{2i},
\end{align*}
and let $Y^{O}_{it} | \cdots$ and $Y^{PM}_{it} | \cdots$ denote the full conditional distributions of missing observations. For the heteroscedastic model, the full conditional distributions for the missing data are
\begin{align*}
Y^{O}_{it} | \cdots &\sim N( \tau^*_{1it} \mu^*_{1it},\tau^*_{1it} ) \\
Y^{PM}_{it} | \cdots &\sim N( \tau^*_{2it} \mu^*_{2it},\tau^*_{2it} ) \nonumber
\end{align*}
with 
\begin{align*}
\tau^*_{1it} &=  \left(\frac{1}{\sigma^2_{1t} } + \sum^{n_{1l}}_{j=1} \frac{\gamma_{1j}^2}{\sigma^2_{1(t+ l_{1j})}} \right)^{-1} \\
\mu^*_{1it} &= \mu_{1it} + \sum_{j=1}^{n_{ 1l } } \frac{ \gamma_{1ij} ( Y^{O}_{i(t + l_{1j} )} - m_{1i(t + l_{1j})} + \gamma_{1ij}  Y^{O}_{it} )}{ \sigma^2_{1(t+l_{1j})} }  \\
\tau^*_{2it} &= \left(\frac{1}{\sigma^2_{2t} } + \sum^{n_{2l}}_{j=1} \frac{\gamma_{2j}^2}{\sigma^2_{2(t+ l_{2j})}} \right)^{-1} \\
\mu^*_{2it} &= \mu_{2it} + \sum_{j=1}^{n_{ 2l } } \frac{ \gamma_{2ij} ( Y^{PM}_{i(t + l_{2j} )} - m_{2i(t + l_{2j} )} + \gamma_{2ij} Y^{PM}_{it} )}{ \sigma^2_{2(t+l_{2j} )} },
\end{align*}
where $l_{1j}$ is the $j^\text{th}$ lag for ozone with coefficient $\gamma_{1ij}$ and $l_{2j}$ is the $j^\text{th}$ lag for $\text{PM}_{10}$  with coefficient $\gamma_{1ij}$. The imputation method for the homoscedastic model is a special case of the heteroscedastic model.

\end{document}